\documentclass[useAMS, usenatbib]{mn2e}
\usepackage{amsmath}
\usepackage[dvips]{graphicx}
\usepackage{graphicx, epsfig}
\usepackage{graphics}

\newcommand{\beq}{\begin{eqnarray}}
\newcommand{\eeq}{\end{eqnarray}}

\newcommand{\apj}{ApJ}
\newcommand{\apjs}{ApJS}

\newcommand{\mnras}{MNRAS}

\newcommand{\nat}{Nature}

\title[Predictability in SAMs]
      {Predictability in Semi-Analytic Models of Galaxy Formation}
\author[J.E. Forero-Romero]
       {Jaime E. Forero-Romero$^{1,2}$\thanks{E-mail:jforero@aip.de}\\
$^{1}$Astrophysikalisches Institut Potsdam, an der Sternwarte 16, D-14482
Potsdam, Germany\\ 
$^{2}$Universit\'e Claude Bernard Lyon 1, CNRS UMR 5574, ENS Lyon, Centre de
Recherche Astrophysique de Lyon, \\Observatoire de Lyon,
9 Avenue Charles Andr\'e, 69561 St-Genis-Laval Cedex, France}
\begin{document}


\pagerange{\pageref{firstpage}--\pageref{lastpage}} 

\maketitle
\label{firstpage}

\begin{abstract}
We propose a general framework to scrutinize the performance of semi-analytic codes of galaxy
formation. The approach is based on the analysis of
the outputs from the model after a series of perturbations in the input
parameters controlling the baryonic physics. The perturbations
are chosen in a way that they do not change the results in the luminosity
function or mass function of the galaxy population. 
 
We apply this approach on a particular semi-analytic model called GalICS. We chose to
perturb the parameters controlling the efficiency of star formation and the
efficiency of supernova feedback. We keep track of the baryonic and observable
properties of the central galaxies in a sample of dark matter halos with
masses ranging from $10^{10}$ M$_\odot$ to $10^{13}$ M$_\odot$.

We find very different responses depending on the halo mass. For small dark
matter halos its central galaxy responds in a highly predictable way to small
perturbation in the star formation and feedback efficiency. For massive dark
matter halos, minor perturbations in the input parameters can induce large
fluctuations on the properties of its central galaxy, at least $\sim 0.1$ in
$B-V$ color or $\sim 0.5$ mag in $U$ or $r$ filter, in a seemingly random
fashion. We quantify this behavior through an objective scalar function we
call predictability.  

We argue that finding the origin of this behavior needs additional information from
other approximations and different semi-analytic codes. Furthermore, the implementation
of an scalar objective function, such as the predictability, opens the door to
quantitative benchmarking of semi-analytic codes based on its numerical
performance.

\end{abstract}


\section{Introduction}
Hierarchical aggregation seems to be at the heart of galaxy evolution. In a
cold dark matter universe, as depicted by numerical  simulations, its
structure grows through subsequent mergers and zero fragmentations. The growth
and evolution of galaxies, which are thought to use dark matter as
scaffolding, is channeled through this hierarchical aggregation, at
least for the most massive structures \citep{LSS_SFW}.

Notwithstanding all the complexity in the process of galaxy formation and
evolution, galaxies still are the most basic population unit in the
description of large scale structure in the Universe. And still nowadays much
work is being invested in galaxy formation to disentangle the influence of the
hierarchical context setup by dark matter from the secular baryonic
processes on small scales. 

Tackling the problem theoretically implies numerical experiments
following large structure dynamics and, at the same time, a
description of baryonic processes such hydrodynamics and radiative
cooling. This is still very challenging in the usual 
numerical approach that discretizes space and time,  and try to solve a
relevant set of equations  to capture the physics
\citep{2002Sci...295...93A,2006AIPC..878....3G}. From the computational point
of view it involves achieving an effective resolution spanning at least $5$
orders of magnitude in mass, length and time \citep{2007arXiv0705.1556N}. 

To overcome this barrier the semi-analytic model (hereafter SAM) approach proposes to
describe first the non-linear clustering of dark matter on large scales, and
describe later the small scale baryonic physics through 
analytical prescriptions. The connection between the two scales is provided through
the dark matter halo, which is the most basic unit of non-linear dark matter
structure (see \cite{2006RPPh...69.3101B} and references therein). 

The non linear clustering of dark matter is described through a merger tree,
representing the merging history of a given dark matter halo.   The
construction methods for merger trees can vary, ranging from Monte-Carlo
realizations based on theoretical estimates \citep{1999MNRAS.305....1S} to the
numerical based on N-body simulations \citep{2006MNRAS.365...11C}, including also hybrid
approaches mixing numerical and analytical techniques
\citep{2002MNRAS.333..623T}.

The different analytic implementations of baryonic process span a wide
range of philosophical approaches, physical concepts and numerical
implementations. Most of them, nonetheless, constructed from observed
correlations in our local patch of Universe.

Regardless of the details of these models, what is general to
all of them is the underlying merger trees structure complementary with
analytic recipes describing the growth of baryonic structure inside the merger
trees. The ignorance respect to the physics included in the analytic recipes
is usually represented by scalars, which in most of the cases represent
efficiencies of physical processes. This implies that a given realization of a
semi-analytic run is completely determined by the dark matter input and the
baryonic parameters in the simulation.

Most of the work during
the last decade was invested in adding and exploring the effect of these
parameters on average quantities, especially the luminosity function. The
generic parameters to be used have been more or less settled, and most of the models have
achieved a good level of internal consistency by reproducing some key
observational features. The confidence on the consistency that can be achieved, and the ease to perform
a semi-analytic run, have empowered the modelers to select subsamples and make
studies about the most massive galaxies or correlations among populations
\citep{2007MNRAS.375....2D,2007arXiv0709.3933H}.

As the complexity and interest in semi analytic techniques grow, two relevant
issues must  be addressed in more detail.  
First, the issue of error  propagation from the incertitude in the
input parameters, a factor that might be  important in a hierarchical Universe,
where the amplification of small initial errors might be important for the most massive and
hierarchical objects. Second,
the development of objective ways to compare different types of complexity in
semi analytic models. This could allow, for instance, the implementation of simple
tests with an objective scalar function to measure the
model performance.  

The objective of this paper is two-fold:
\begin{itemize}
\item{Propose a methodology to weigh the role of secular
    baryonic processes in the context of SAMs.}
\item{Propose an objective scalar function that captures the biases and
  general behavior of semi-analytic models regardless of its detailed implementation.}
\end{itemize}

These two objectives are a result of the same perturbative approach we
advocate in this paper. This approach is based on the fact that the only
objective information we have to describe the results of a semi-analytic run are
the input parameters of the model. In the perturbative
approach, we perform semi-analytic runs in the neighborhood of some scalar
parameters. This will allow us, as we will show, to get an idea
about the limits of our semi-analytic model.

This paper is structured as follows. In Section \ref{sec:SAM} we describe the
structure common to all SAMs and from that point we introduce the
concept of perturbations in a semi-analytic model. In Section 3 we introduce
the setup for the perturbation experiment of our SAM. We
describe in Section 4 the two most relevant qualitative features of the
experiment results. We select one of this qualitative results to make a
detailed quantitative analysis with three different indices, these results are
shown in Section 5. We discuss our results in Section 6.

\section{Semi-Analytic Models}
\label{sec:SAM}
\subsection{Common features}
Semi-analytic models exploit the fact that there are two very
different physical scales involved in the process of galaxy formation and
evolution. On large scales dark matter and gravity are dominant, while  on smaller
scales complex radiative processes are central to the development of galactic
sized structures \citep{1999MNRAS.310.1087S,galicsI,2003MNRAS.343..367B,2006MNRAS.365...11C,MORGANA}.

Inside semi-analytic models all the non-linear dark matter dynamic is
described through the merger tree, which represent the process of
successive mergers building a dark matter halo. On top
of this merger tree, all the complex baryonic physics are implemented
through analytic prescriptions derived in most part from observations.

The baryonic processes, in the end, are controlled by a set of scalars,
which represent most of the time either an efficiency or a threshold value.
From the pure functional point of view, all the baryonic properties
$\mathcal{B}$ of a dark matter halo $\mathcal{H}$ are a function
of its merger tree $\mathcal{T}$ and the set of scalar parameters
controlling the model $\{\lambda_1\ldots\lambda_N\}$.

\begin{equation}
  \mathcal{B} = \mathcal{B}(\mathcal{T}, \lambda_1,\ldots,\lambda_N).
\label{base}
\end{equation}

Furthermore, during a semi-analytic run, the set of parameter $\{\lambda_1\ldots\lambda_N\}$ is
fixed to be the same for all the halos, all the time. Thus, the trees and the
scalar values completely define the outputs.


From the perspective of disentangling the role of different physical
elements in the process of galaxy formation, the approach commonly followed is
the exploration through a set of different 
values for the $\{\lambda_{i}\}$ parameters, taking as a gauge the
reproduction of the luminosity function of diverse galaxy populations. This
coarse exploration of parameter space have been done until a minimum internal
consistency is achieved, a decision based on the success of reproducing a wide set of
observational constraints. 

Nowadays more and more results of semi-analytic
models are being used in the predictive sense, selecting subsamples of
galaxies, trying to explain or predict astrophysical quantities of interest
based on the results of a semi-analytic model
\citep{2007MNRAS.375....2D,2007arXiv0709.3933H}. This have been done without
an explicit treatment of the potential biases and complications introduced by
the semi-analytic model itself.

We are interested in understanding in greater detail the behavior and
limits of semi-analytic methods, regardless of its detailed implementation
across different codes. Our attempt to deal
with the complexity across semi-analytic models, is based on the basic conceptual approach in
Eq.\ref{base}, which is the only structural information we have about how
semi-analytic models work. 

\subsection{Perturbing the model}

We intend to measure the effect of perturbations in the model
\begin{equation}
  \mathcal{B}^{\prime} = \mathcal{B}(\mathcal{T},
  \lambda_1+\delta\lambda_1,\ldots,\lambda_N+\delta\lambda_N), 
\label{base-pertirbed}
\end{equation}

where the magnitudes of the perturbations $\delta\lambda_i$ are constrained in
such a way that its effect is not significant on the mean population quantities such as
the luminosity function, meaning that we are not breaking the broad
consistency of the model.

The main objective is measuring the consequences of these
perturbation and to use them as a gauge of the model's numerical performance, but
also to see how and where are going to emerge
the consequences of the perturbations.

If we intend to explore the neighborhood of a given set of parameters
$\{\lambda_i\}$ by making runs around $\{\pm \delta\lambda_{i}\}$, this implies
performing $2^{N}$ different runs where $N$ is the total number of scalar
parameters controlling the model. If we want to explore the neighborhood
around $m$ different values for each $\lambda_i$, the number of runs becomes
$m^N$.

The number of free parameters in SAMs can be at least $6$. Which
means that we should deal at least with $2^6=64$ different runs to minimally
explore the neighborhood of $\{\lambda_i\}$. Performing all these runs over a
cosmological volume is unfeasible and perhaps not very useful.

The approach we decided to follow in this paper uses two
simplifications. First, we only perform the simulations over a subset of
dark matter halos selected at random in a box of cosmological size. Second, we explore only
the neighborhood of two scalar parameters.

The size of the halo subsample is about $1\%$ of the total number of halos in
the simulated cosmological box, and the parameters we will explore are the star
formation efficiency $\alpha$ and the supernova feedback efficiency $\epsilon$.

\section{Experiment}
The semi-analytic model we use in this paper is a slightly modified
version of that presented in \cite{galicsI}. As we do not want to compare our results with
observations or another models, we briefly review only the elements relevant
for our discussion: the dark matter description and the star formation and
supernovae feedback implementations. 

\subsection{Dark Matter}

The dark matter simulation was performed using cosmological parameters compatible with a
1st year WMAP cosmology \citep{2003ApJS..148..175S} ($\Omega_m$, $\Omega_\Lambda$, $\sigma_8$, $h$) $=$
($0.30$, $0.70$, $0.92$, $0.70$), where the parameters stand for the density of
matter, density of dark energy, amplitude of the mass density fluctuations and
the Hubble constant in units of $100$ km s$^{-1}$ Mpc $^{-1}$. . 
The simulation volume is a cubic box of side $100 h^{-1}$ Mpc with $512^{3}$
dark matter particles, which sets the mass of each particle to  $5.16
\times 10^{8}\ h^{-1}$ M$_\odot$. The simulation was evolved from an initial
redshift $z=32$ down to redshift $z=0$, keeping the particle data for $100$
time-steps.

For each recorded timestep build a halo catalogue using a friends-of-friends
algorithm \citep{FOF} with linking length $b=0.2$. Only the groups with $20$ or more bound particles are
identified as halos. This sets the minimal mass for a dark matter halo to $ 1.03\times
10^{10}\ h^{-1}$ M$_\odot$. These halo catalogues provide the input for the
construction of the merger trees used as input for the semi-analytic model.

\subsection{Star Formation and SN feedback}
The star formation rate is set proportional proportional to the mass
of cold gas, and without any other characteristic time scale we impose
that the rate at which the gas is consumed to form stars is given by
the dynamical time of the disc. This is motivated by the observational
correlations observed by Kennicutt-Schmidt \citep{1998ApJ...498..541K}. 

Hence, in our model, the global star formation rate $\Psi_{\star}$ on galactic scales is
given by the following equation   

\begin{equation}
\Psi_{\star}  = \alpha\frac{m_{gas}}{t_{dyn}}, 
\end{equation}

where $\alpha$ is an efficiency parameter, and $t_{dyn}$ is the
dynamical timescale of the component we are interested (disc or bulge). For $t_{dyn}$
we use the time taken for material at the half-mass radius to reach
either the opposite side of the galaxy (disc) or its center (bulge), and is given by:  

\begin{equation}
t_{dyn}= r_{1/2}\times \pi v ^{-1}, 
\end{equation}

where $v$ is a characteristic velocity in the galaxy component and $r_{1/2}$
is the half mass radius. For discs the
velocity $v$ is equal to the circular velocity of the disc where the material is assumed to
have purely circular orbits. In the case of spheroidal components $v$ is the velocity dispersion,
where we assume the matter in the component has only radial orbits.  

The star formation is triggered if the column density of the gas is greater
that a given threshold constrained by the observations \cite{1998ApJ...498..541K}. By simplicity we
assume that the initial mass function is universal at all redshift and follows
a Kennicutt initial mass function.  

Once stars are formed, the massive stars will explode inside the galaxies
ejecting hot gas and metals in the interstellar medium. The simple
model that we use for this phenomenon is given by the implementation of
\cite{2001MNRAS.324..313S}, where the rate of gas mass loss is written assuming an stationary model 

\begin{equation}
\frac{dm_{out}}{dt} = \Psi_{\star}\times\eta_{SN}\Delta
m_{SN}\times(1+L)\times(1-e^{-R}), 
\end{equation}

where $\Psi_{\star}$ is the star formation rate, $\eta_{SN}$ is the number of supernovae per
unit mass of formed stars (fixed number function of the IMF), $\Delta
m_{SN}$ is mean mass loss of one supernova ($\sim 10$ M$_\odot$) and $(1+L)$
is defined as

\begin{equation}
1+L\equiv\epsilon\frac{m_{gas}}{m_{gal}}, 
\end{equation}

where $m_{gas}$ and $m_{gal}$ are the gas and total mass of the galaxy
component and the parameter $\epsilon$ regulates the efficiency of the
feedback. We also define the porosity of the galaxy component as 

\begin{equation}
R = \alpha \left(\frac{m_{gal}}{m_{gas}}\right)^{1/2}\left(\frac{17.8}{\sigma}\right)^{2.72},
\end{equation}

where $\sigma$ is a typical dispersion velocity in the interstellar medium (in km/s), which we
fix to $10$km/s for disks and to the velocity dispersion for the spheroidal
components. The parameter $\alpha$ is the star formation efficiency. In this
model, usually the ejected amount of gas is of the same order of the mass of
formed stars.

\subsection{Experiment Setup}
From the detected halos at redshift zero we select at random nearly
$600$ of them, which corresponds to about $1\%$ of the total number of
halos in the box. For each halo we have its corresponding merger history, and we are
able to run our galaxy formation code on every individual merger tree\footnote{Actually because of technical reasons the code is run over a bundle
  of merger trees.}. 
 
For each halo we make 320 runs varying two parameters, the star formation
efficiency $\alpha$ and the supernova feedback efficiency $\epsilon$. The
first parameter, $\alpha$, is sampled at 16 points between 0.018 and 0.022,
and the second, $\epsilon$, is sampled at 20 points between 0.18 and 0.20.
From this we define a 2-dimensional representation to keep track of each
run. We setup two coordinate plane. Along the first dimension ($x$-axis) we
vary the star formation efficiency $\alpha$, and along the second dimension ($y$-axis)
we vary the supernova feedback efficiency $\epsilon$. This defines the
$\alpha$-$\epsilon$ plane.

For each run (for a given halo and for given point in the $\alpha$-$\epsilon$
plane) we select the central galaxy in the halo, which is the only
galaxy with a clear identity in a hierarchical paradigm. For each central
galaxy we track six physical properties: total mass (gas and
stars), mass of stars , bolometric luminosity, absolute magnitude in the
SDSS$_U$, SDSS$_r$ filters and the $B-V$ color. We will
refer to the values of a given galactic property, for a given galaxy, over the
$\alpha$-$\epsilon$ plane as a \emph{landscape}.

\section{Qualitative Results}

\begin{figure*}
\begin{center}
\includegraphics[scale=0.45]{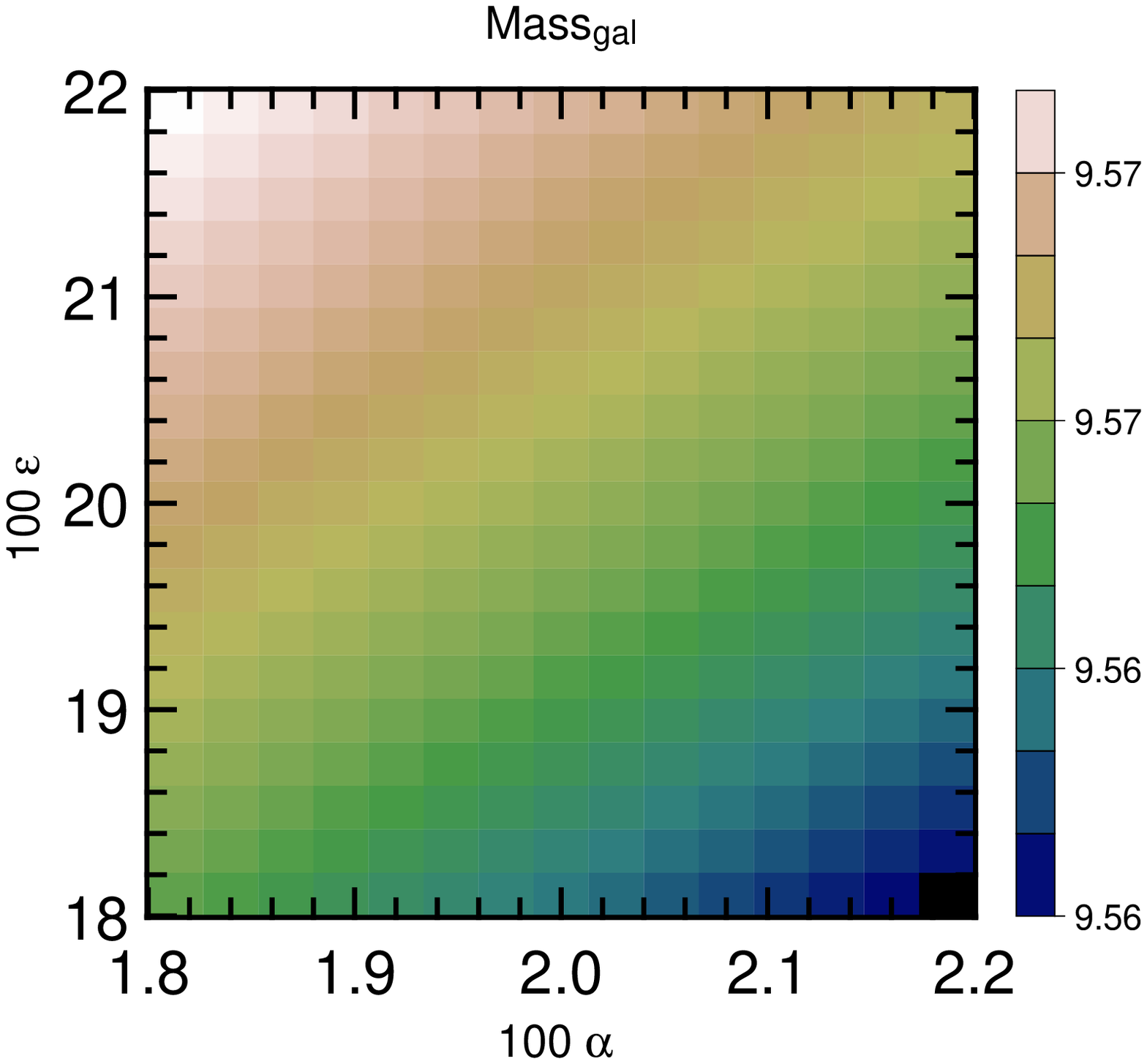}
\includegraphics[scale=0.45]{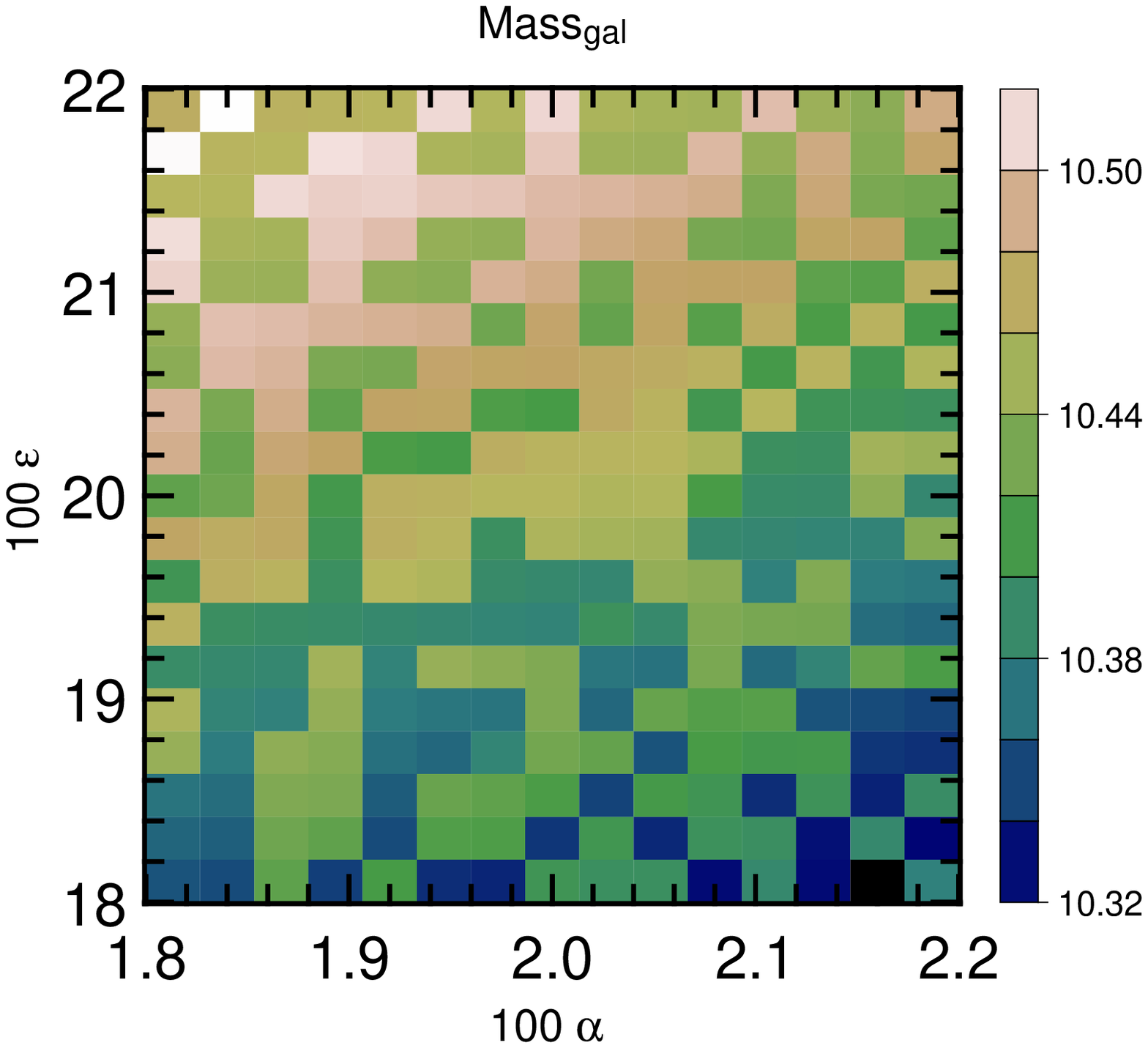}
\end{center}
\begin{center}
\includegraphics[scale=0.45]{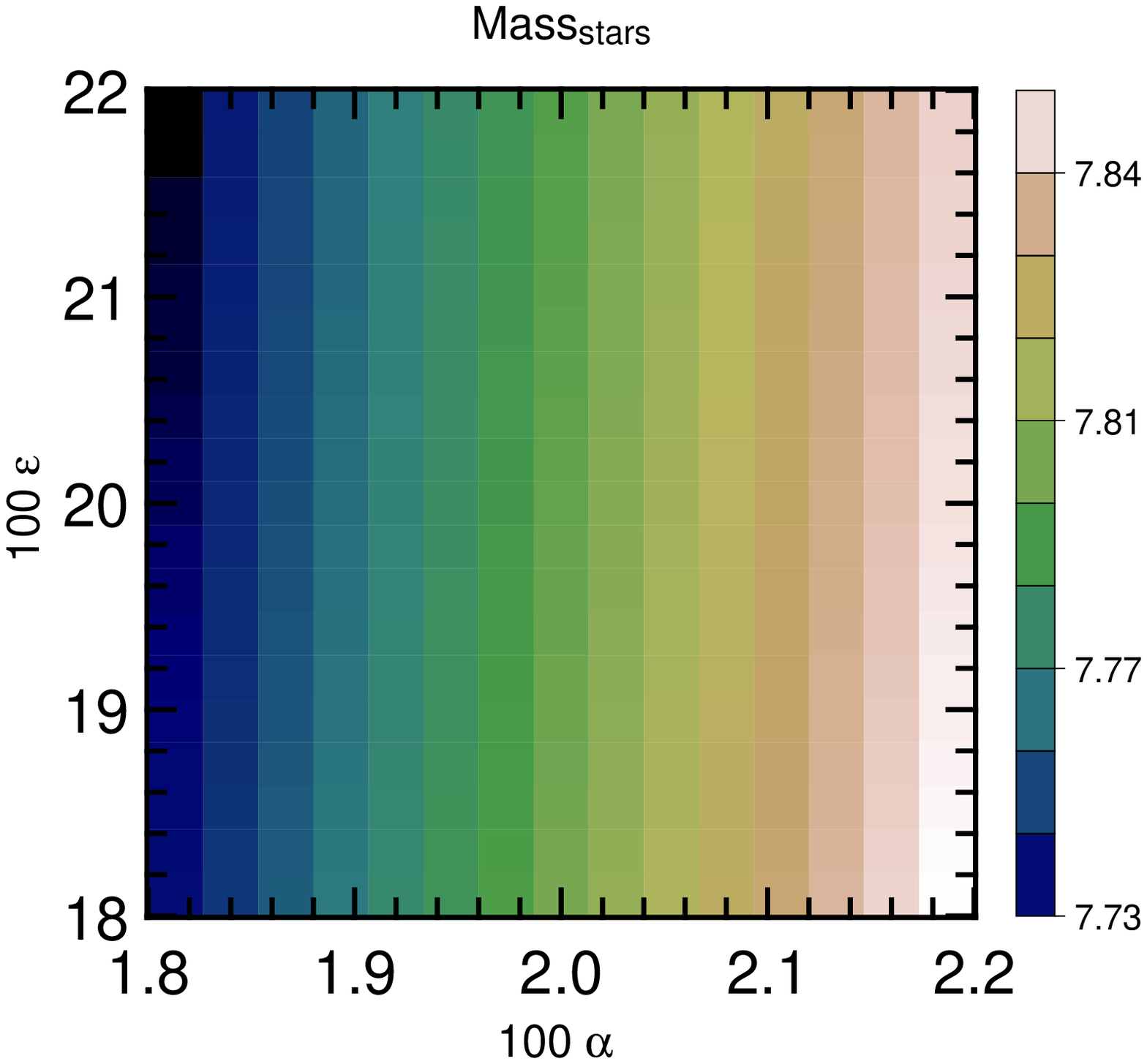}
\includegraphics[scale=0.45]{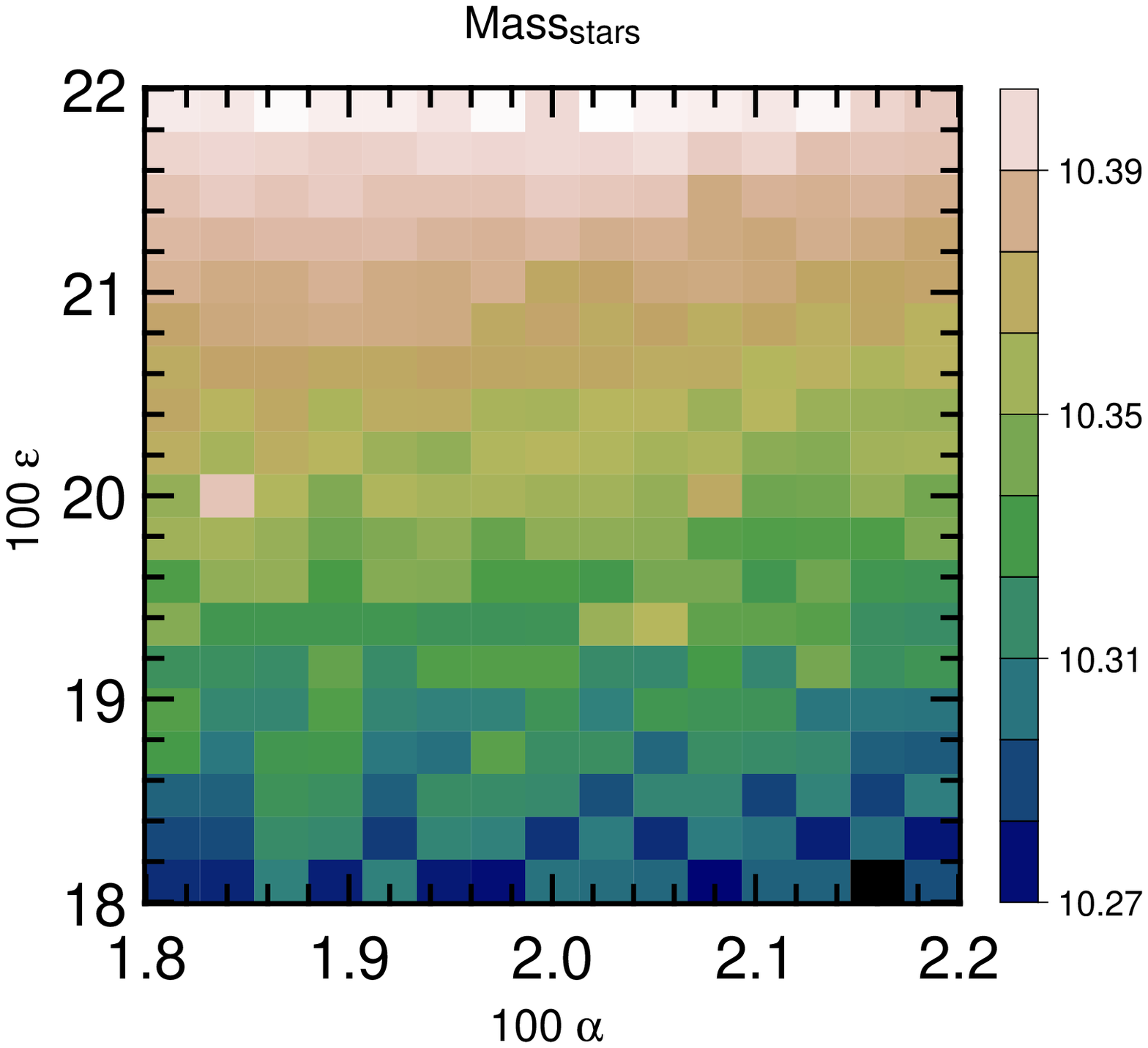}
\end{center}
\begin{center}
\includegraphics[scale=0.45]{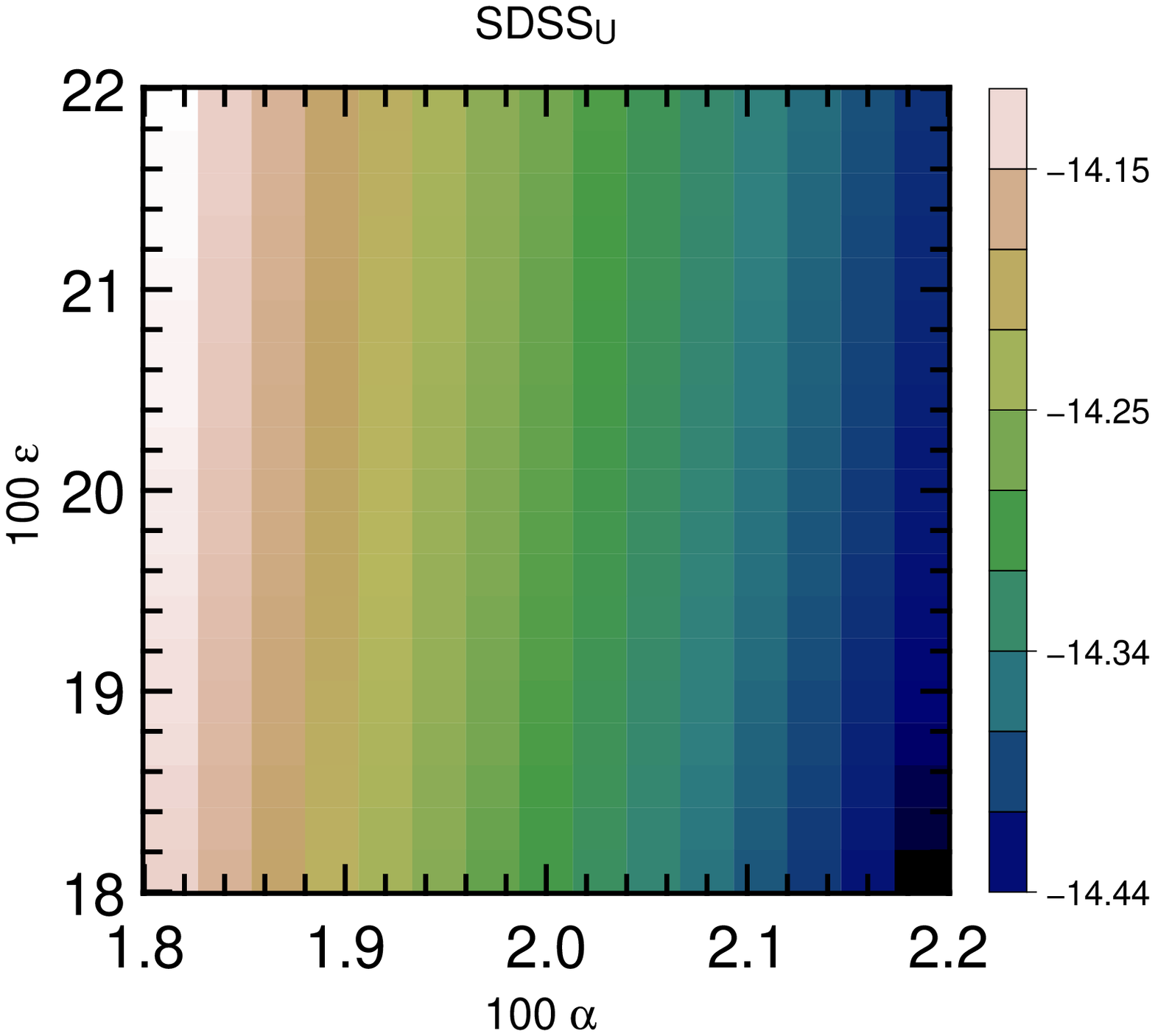}
\includegraphics[scale=0.45]{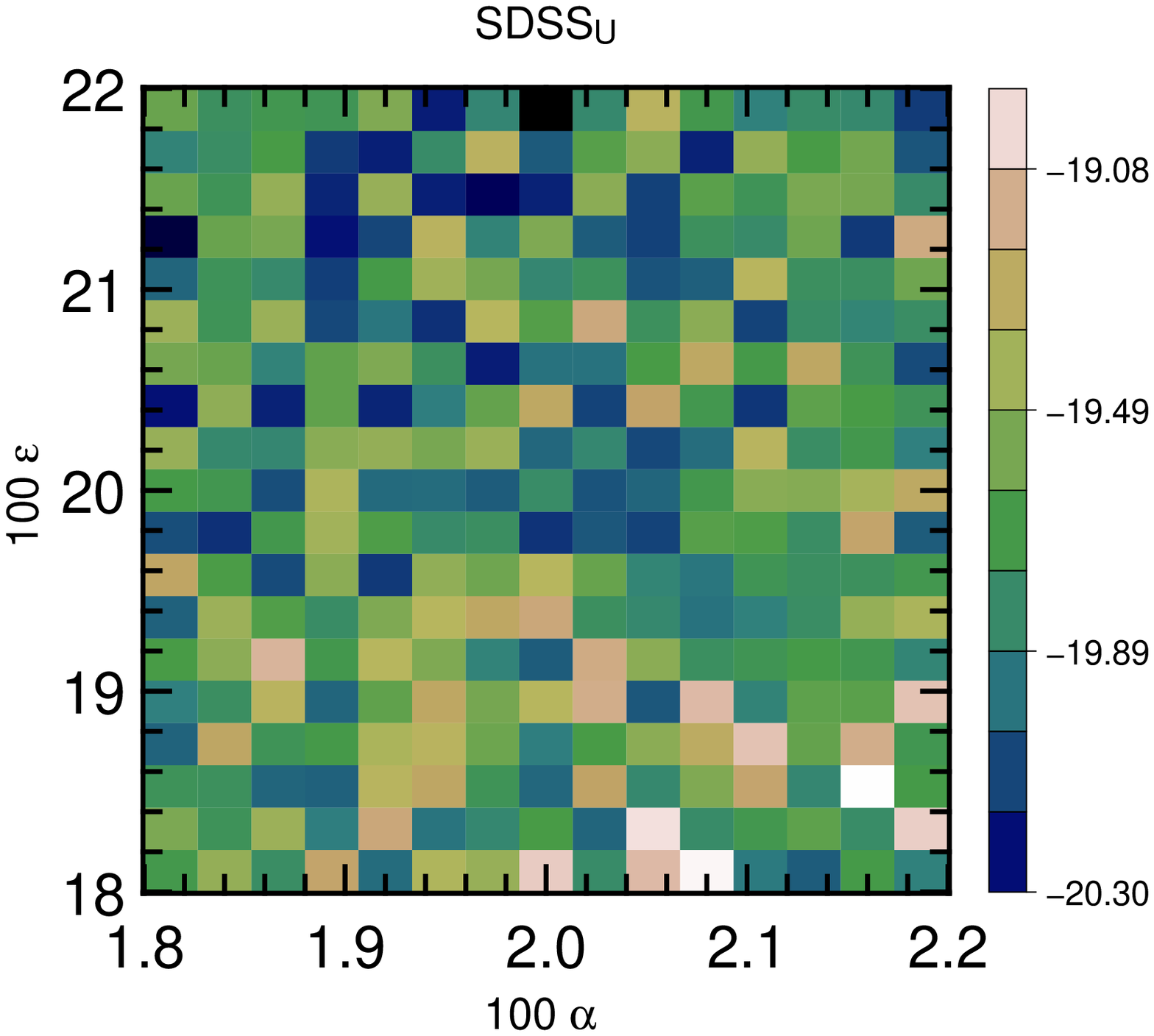}
\end{center}
\caption{\label{meanhalo} Results for the sampling of the star formation
  efficiency $\alpha$ and the supernova feedback efficiency $\epsilon$. 
  Each panel represents the results of some galactic property in the central
  galaxy of a dark matter halo. On the left, the results correspond to the central galaxy in a 
  halo of mass $\sim 10^{10}$ M$_\odot$. On the right, to a more massive halo
  of $\sim 10^{12}$  M$_\odot$. For each galaxy we  show the results concerning the total galaxy
  mass (upper panels), the stellar mass (middle panels) and the
  magnitude in the SDSS$_U$ band (lower panels). Every small square in each
  panel shows the result at redshift $z=0$ for the run with the corresponding
  value of $\alpha$ and $\epsilon$. The results for the low mass halo are
  predictable, for the high mass halo they are almost random. Note that for
  instance in the case of the SDSS$_U$ filter the values fluctuate over a
  range of $\sim 1.4$ mag. The bulk of our paper is devoted to the
  quantification of this behavior as a function of halo mass.} 
\end{figure*}

\begin{figure*}
\begin{center}
\includegraphics[scale=0.25]{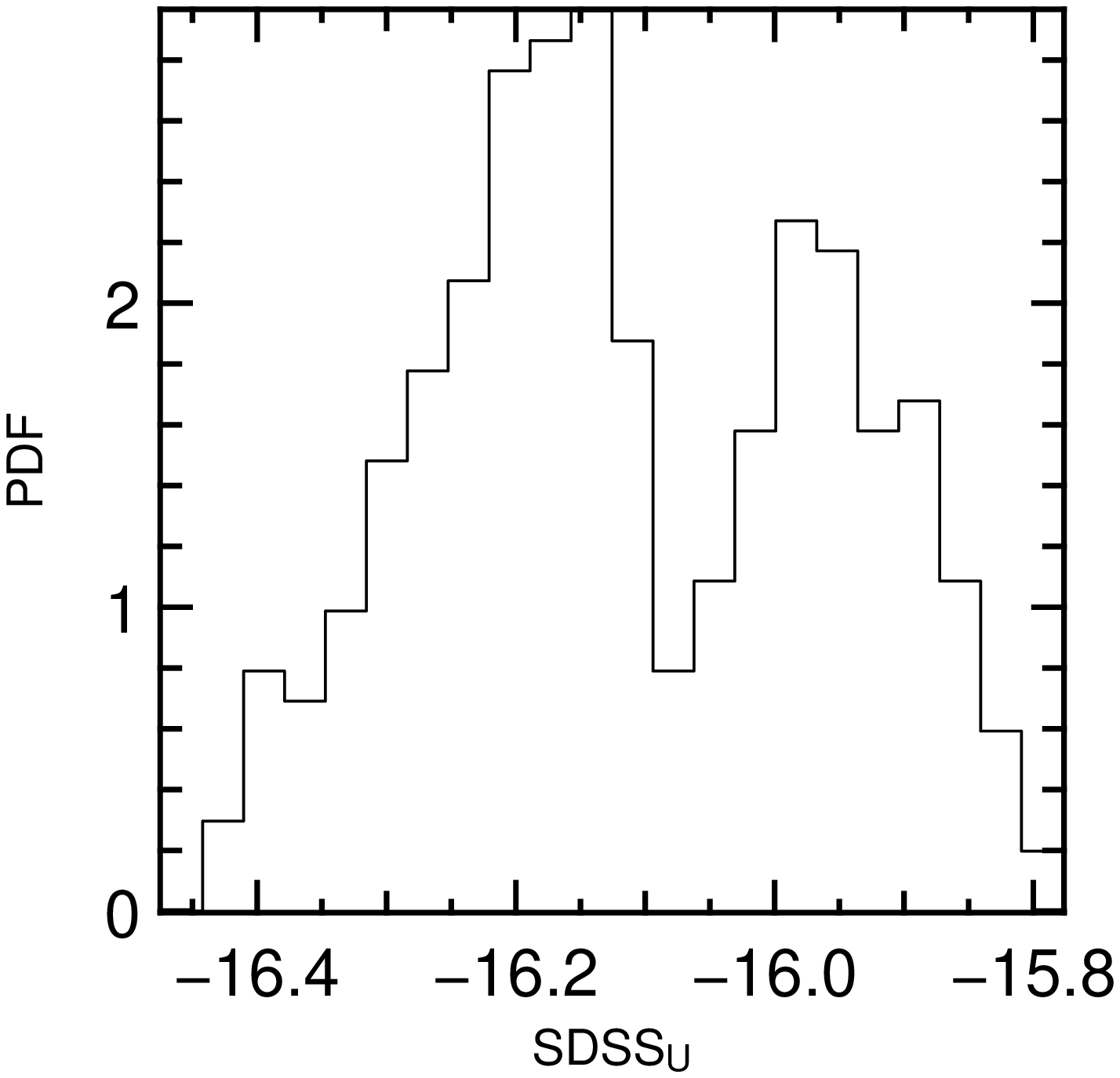}\hspace{2mm}
\includegraphics[scale=0.25]{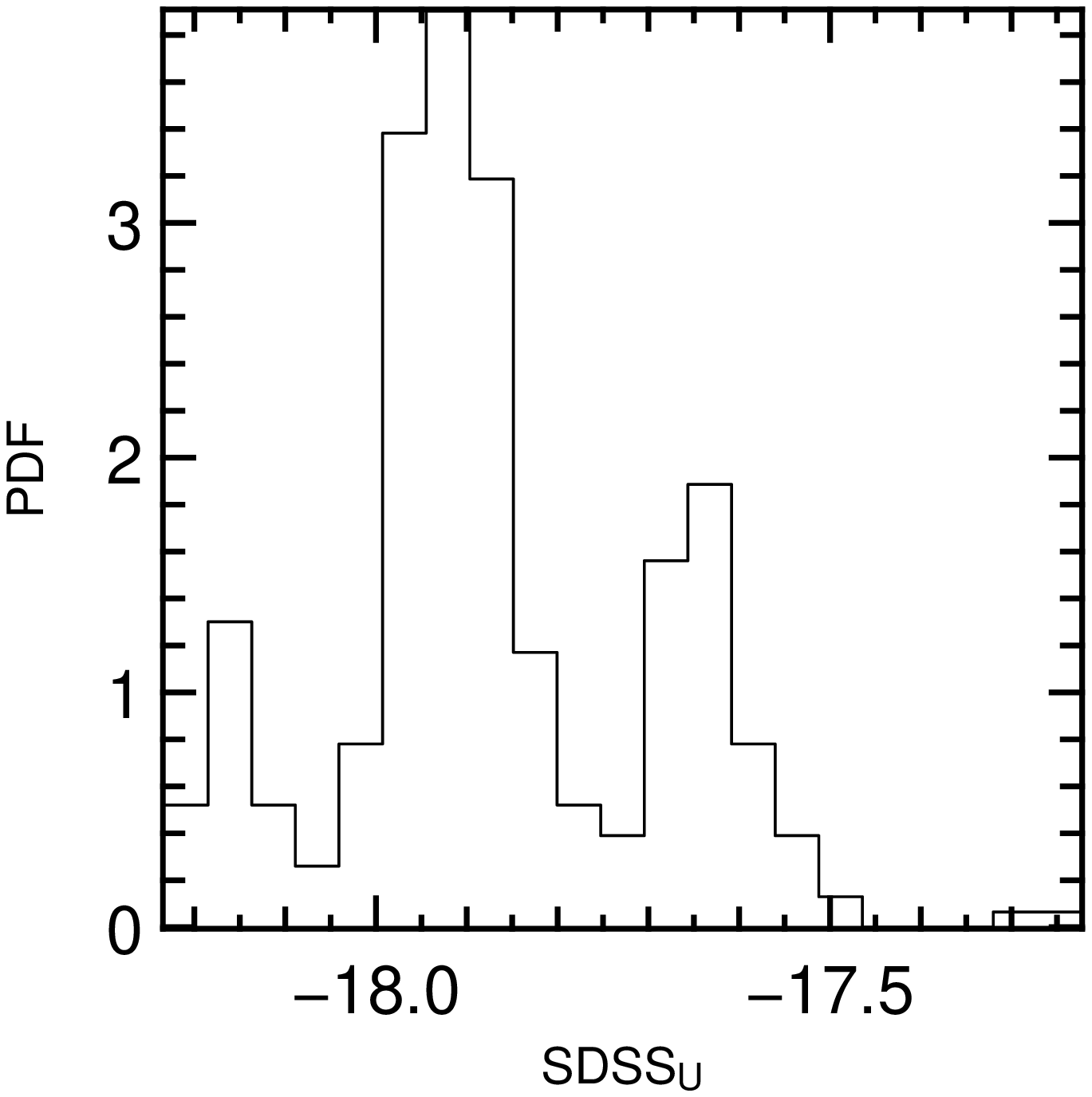}\hspace{2mm}
\includegraphics[scale=0.25]{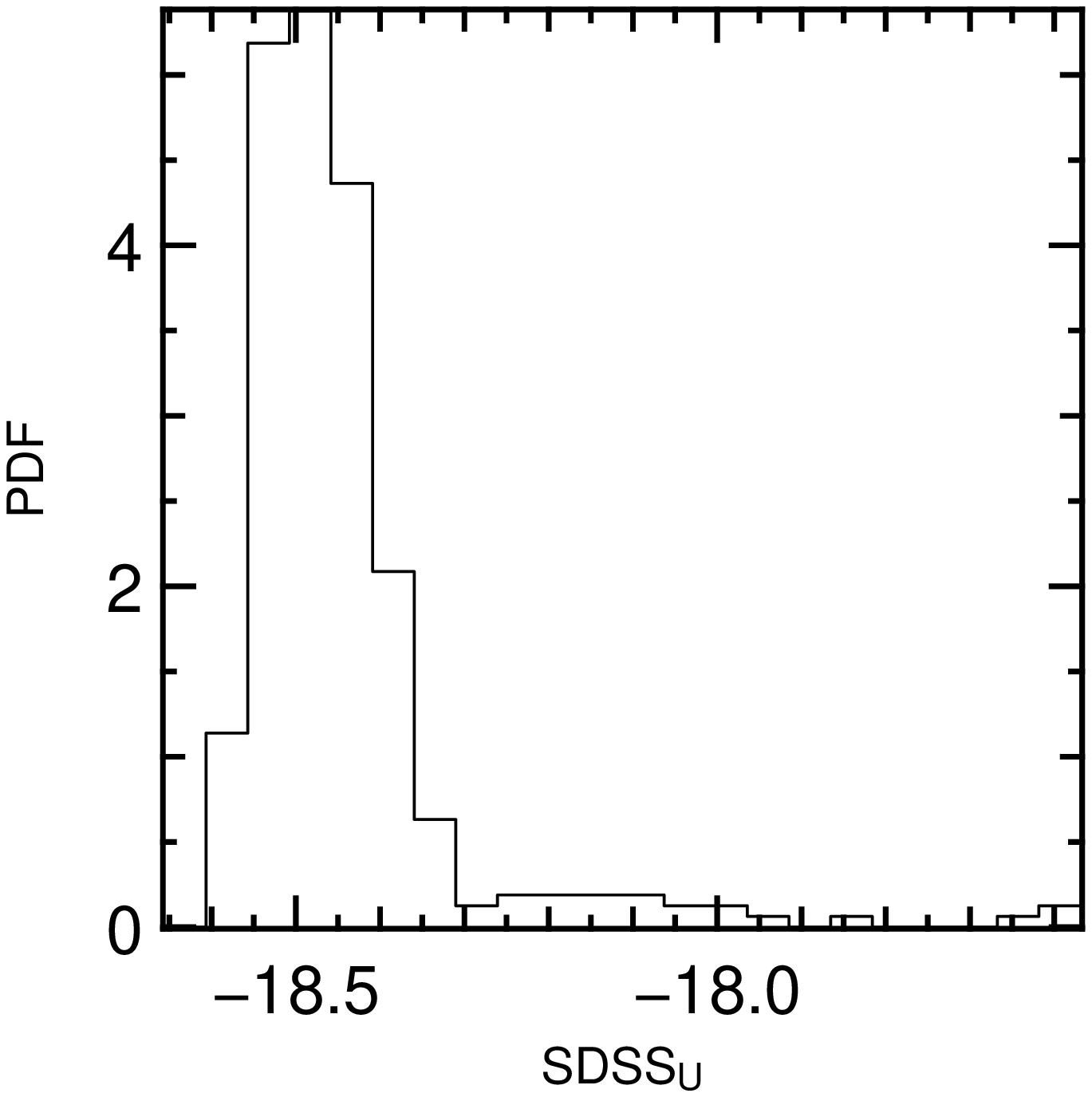}\hspace{2mm}
\includegraphics[scale=0.25]{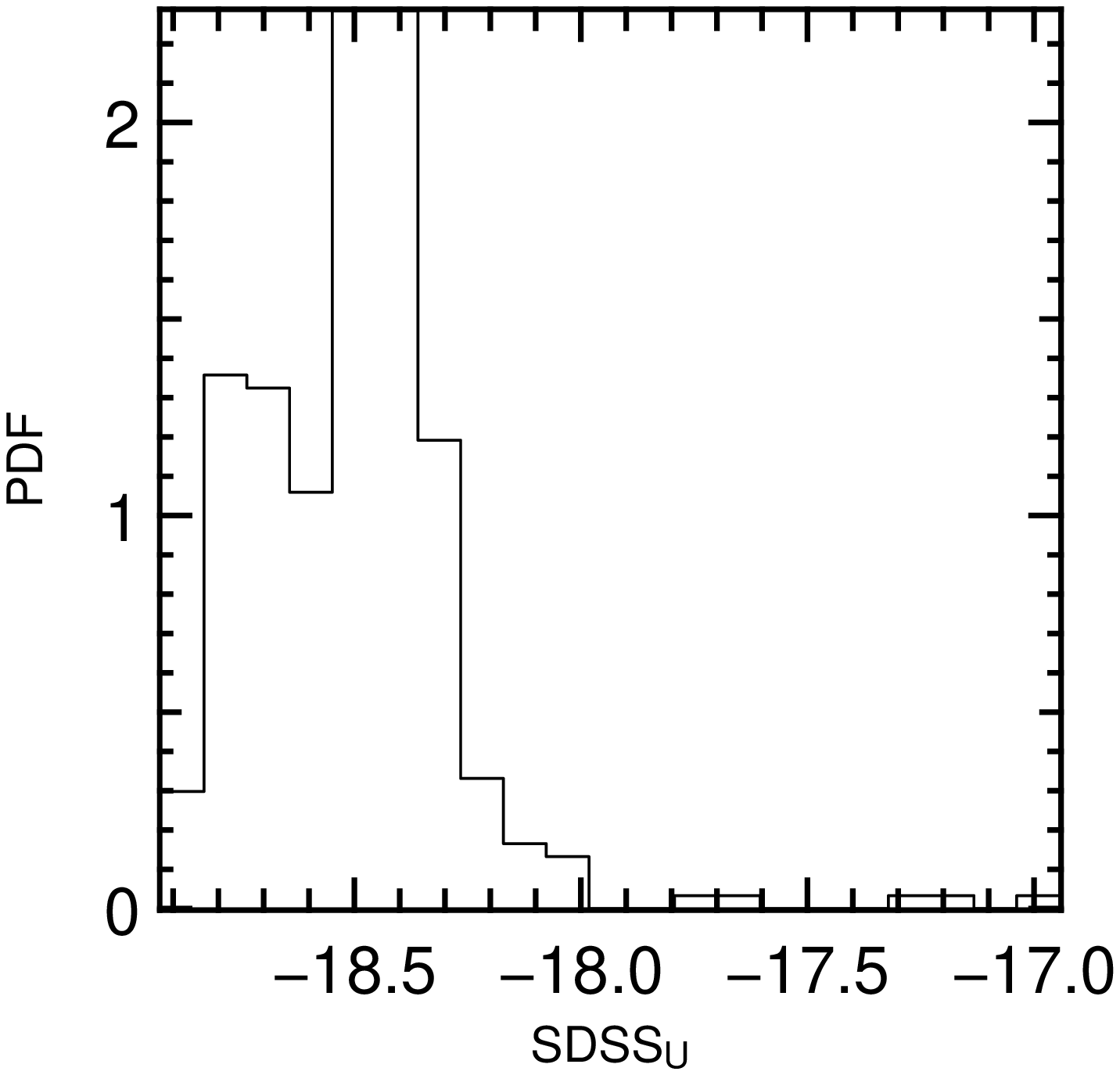}\hspace{2mm}
\end{center}
\begin{center}
\includegraphics[scale=0.25]{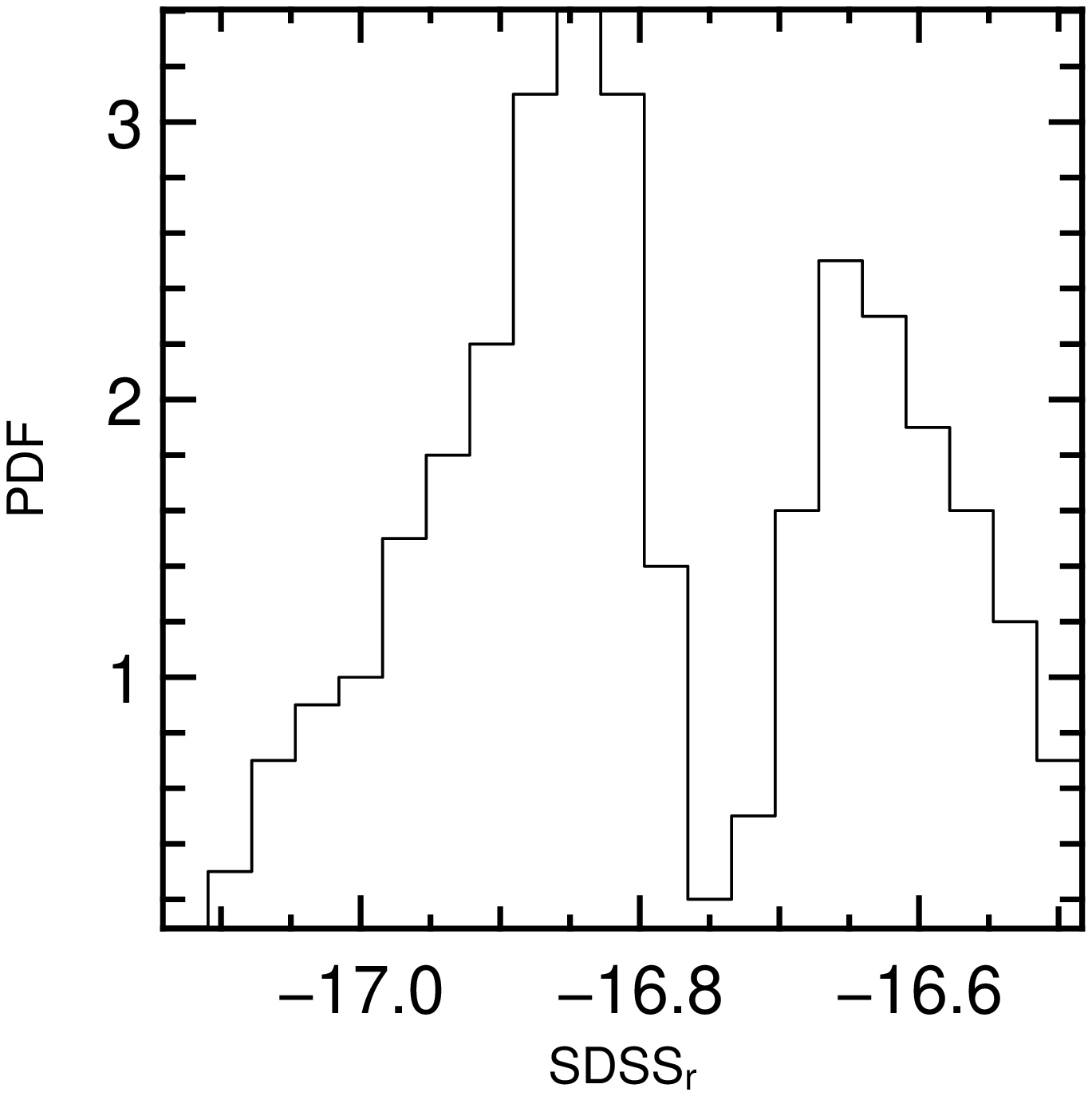}\hspace{2mm}
\includegraphics[scale=0.25]{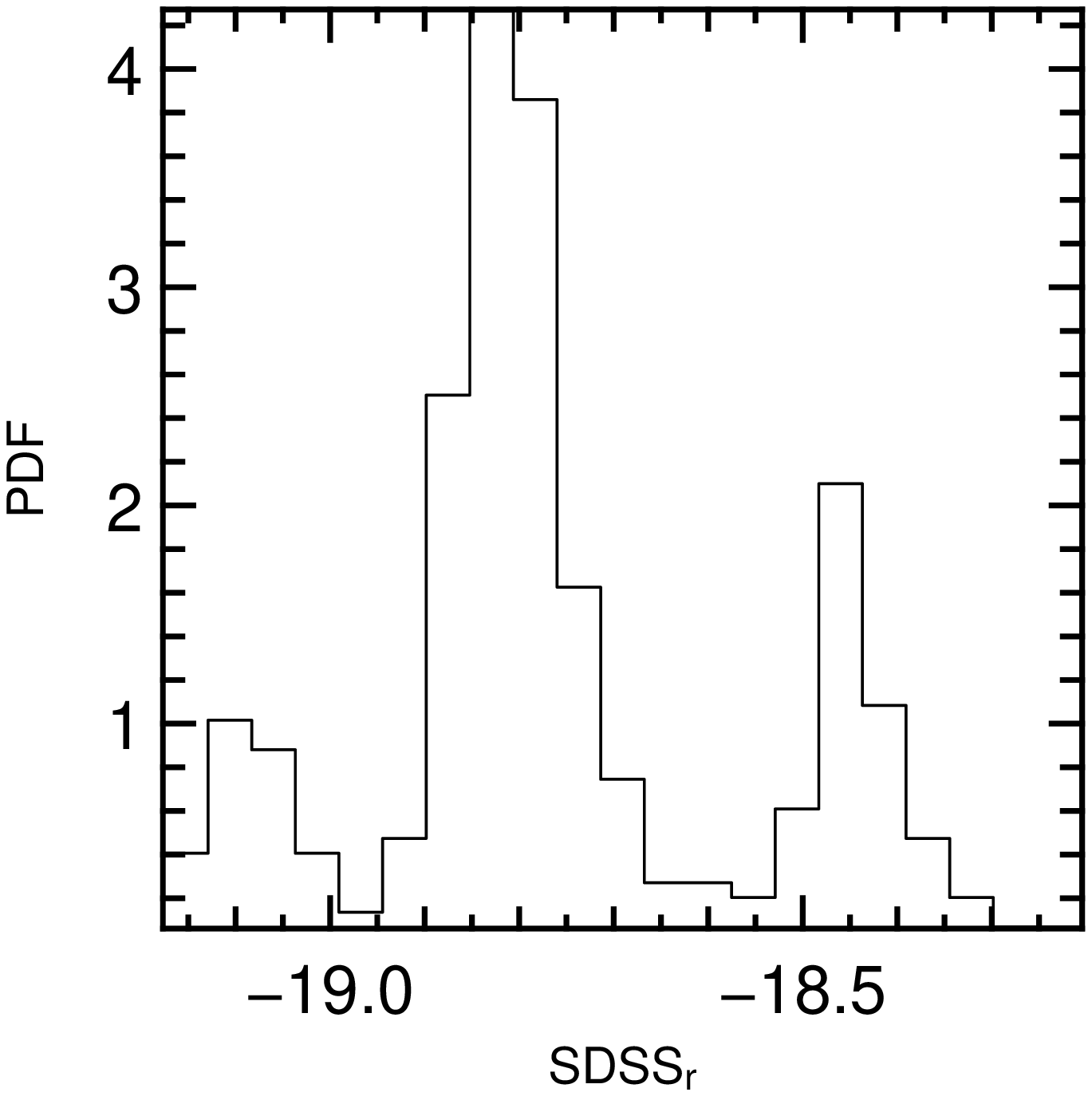}\hspace{2mm}
\includegraphics[scale=0.25]{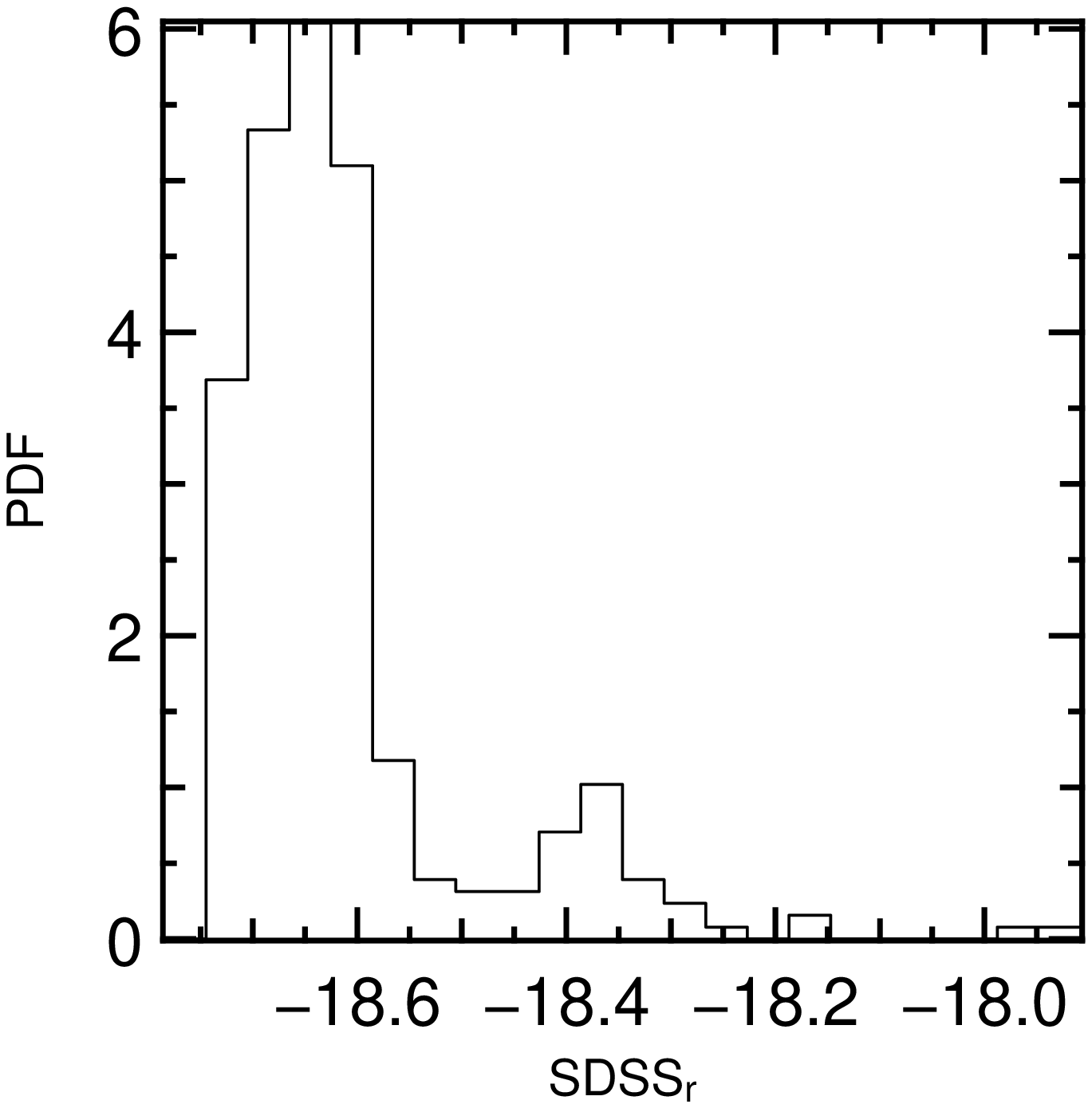}\hspace{2mm}
\includegraphics[scale=0.25]{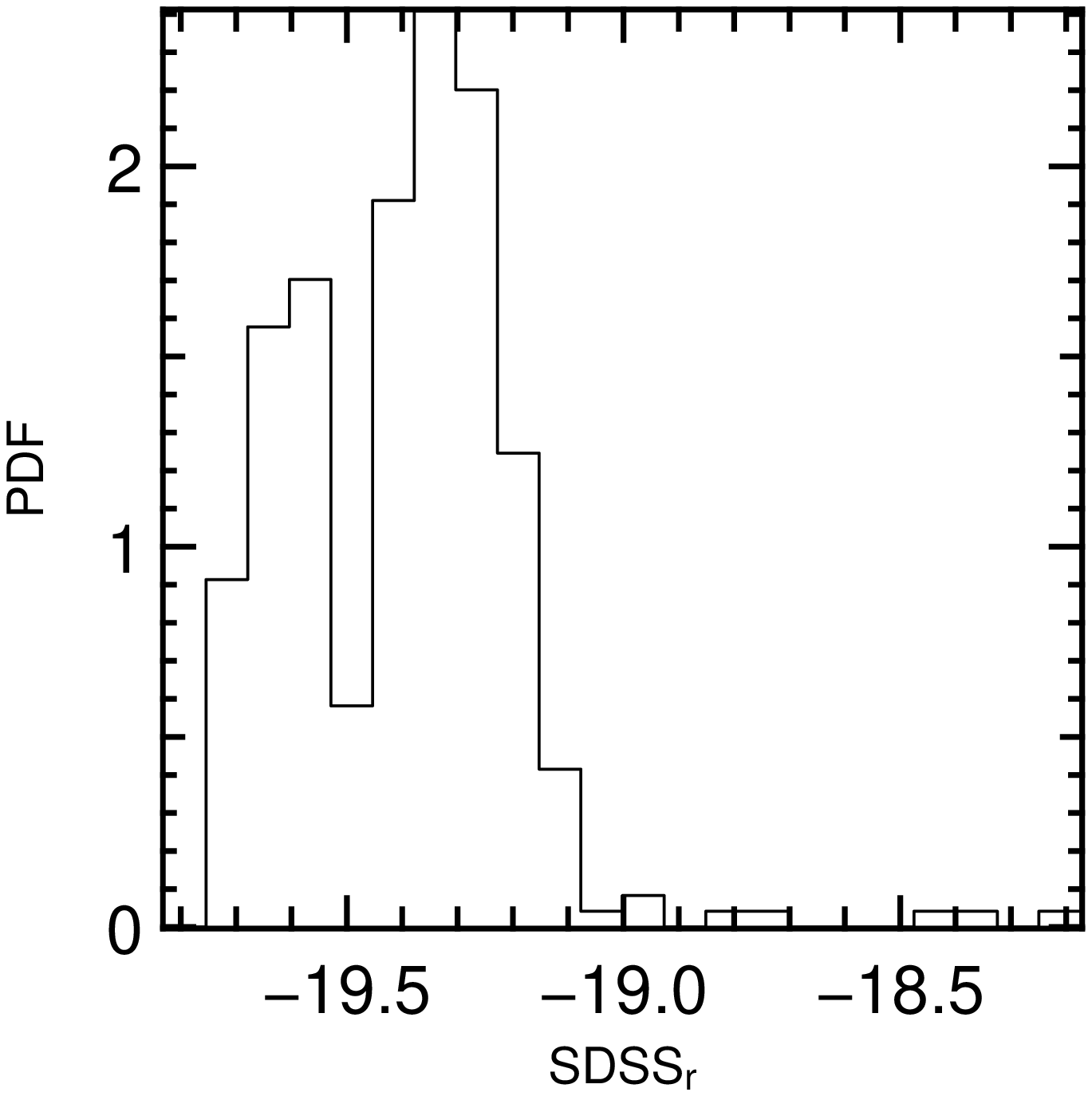}\hspace{2mm}
\end{center}
\caption{\label{histohalo} Histograms (normalized to add up to unity) of the
  values over four randomly selected the landscapes. From left to right the mass of the host dark
  matter halo increases. Upper row  shows the results for the SDSS$_U$
  magnitudes. Lower row: SDSS$_r$   magnitudes. This illustrates another
  qualitative feature of the landscapes, namely that sometimes they are
  bimodal, for instance in the upper-left panel. We do not try to quantify
  this behavior in the paper.} 
\end{figure*}

We present in Fig.\ref{meanhalo} the landscapes for the total galactic
mass, stellar mass and SDSS$_U$ absolute magnitude for two
galaxies, each column representing one galaxy. Qualitatively speaking, we can
spot a striking difference in this figure. 

The left column in Fig.\ref{meanhalo} presents the results for a central galaxy in  halo of mass
$\sim 10^{11}$M$_{\odot}$. Its growth process  have been dominated by what we
call smooth accretion, meaning that at our working resolution this halo has
not suffered any major merger. The predicted properties vary
smoothly over the $\alpha$-$\epsilon$ plane. 

On the right column in the same figure, we show the same properties for the
central galaxy in a halo of mass $\sim 10^{13}$M$_{\odot}$. In this case the
values over the landscape do not follow any pattern. The biggest difference
with respect to the previous case is that the halo growth cannot be described
by pure accretion, but  through repeated mergers.

A second qualitative feature is the emerging bimodality for some
landscapes. It is visible in the upper right panel in Fig.\ref{meanhalo},
where it seems that the values over the landscape are oscillating back and
forth between  two planes. To illustrate better this effect we have
constructed the histograms for two kind of landscapes (SDSS$_r$ and
SDSS$_U$ magnitudes) for four different halo masses. The results are shown in 
Fig.\ref{histohalo}, which shows how the landscapes are not necessarily
unimodal. By visual inspection of half of the landscapes for the total mass, bolometric
luminosity and SDSS$_r$, we can report that the non-unimodality is a recurrent
landscape feature.

For the rest of the paper we will be concerned with a quantification of the
first result, which showed an apparent randomness for the central galaxies in strongly
hierarchical halos. We will use three different indicators. 

First, we will define a scalar function called predictability, $P$, for a given
galactic property over the $\alpha$-$\epsilon$ plane \citep{predictpaper}. The predictability will
be almost one for the low mass case, and zero (or even negative) for the
case of the massive halo. 

The second method of quantification is based on the predictability and the
variance over the landscapes. We will calculate a predictability-weighted
variance, which is intended to represent a quantitative estimation of the
variations we can expect in a galactic property after performing a minimal
perturbation $\delta\alpha$-$\delta\epsilon$. 

The last method of quantification compares the variance over the landscapes
with the variance over a subsample of galaxies hosted by halos of similar mass
inside the full cosmological box.

\section{Quantitative Results}
\subsection{Predictability}

We present the first part of the qualitatively results using a scalar function
we call predictability.  First, we sketch out the general idea behind its definition. 

We place ourselves on the $\alpha$-$\epsilon$ plane, and we want to predict the value of
some galactic quantity at the point we are standing, we also intend to use the
information available in the neighborhood. We have the values of the quantity
we want to measure for the four nearest neighbors in the $\alpha$ -$\epsilon$
plane. We make a guess for that value by averaging these values, and at the
same time we perform the measurement.    

We have now two different values at the point in the $\alpha$ -$\epsilon$ plane,
one is predicted and the other is measured. If the squared difference between
these two values is small for each point in the plane, we can be sure that
we are over a smooth landscape. If the squared differences over the plane are
big, the landscape is not so smooth. The predictability is a measure based on
these squared differences.

 In practice, we use a discretization of the plane $\alpha$-$\epsilon$, and we
 construct two different scalar fields over that plane. The first 
 corresponds to the field measured in the numerical runs, noted $L$. The
 second is a predicted version, noted $L^{\prime}$. 
 
 The values of $L^{\prime}(\alpha_{i},\epsilon_{j})$ are calculated from the
 neighboring points in $L(\alpha,\epsilon)$, as follows 
 \begin{multline}
   L^{\prime}(\alpha_{i},\epsilon_{j}) =
   \frac{1}{4}[L(\alpha_{i+1},\epsilon_{j})+ \\
   L(\alpha_{i-1},\epsilon_{j})+L(\alpha_{i},\epsilon_{j+1})+L(\alpha_{i},\epsilon_{j-1})].
\end{multline}

We construct now the following quantity 
 
\begin{equation}
Q^2 = \frac{1}{N}\sum_{i,j} [L^{\prime}(\alpha_{i},\epsilon_{j})-L(\alpha_{i},\epsilon_{j})]^2,
\end{equation}

where $N$ is the total number of points in the plane $\alpha$-
$\epsilon$. This quantity help us to define the predictability 

\begin{equation}
P = 1 - \frac{Q^2}{\sigma^2},
\label{predigo}
\end{equation}

with $\sigma^{2}$ as the variance of the landscape

\begin{equation}
\sigma^2 = \frac{1}{N}\sum_{i,j} [L(\alpha_{i},\epsilon_{j})-\bar{L}]^2.
\label{land_variance}
\end{equation}

\begin{figure*}
\begin{center}
\includegraphics[scale=0.33]{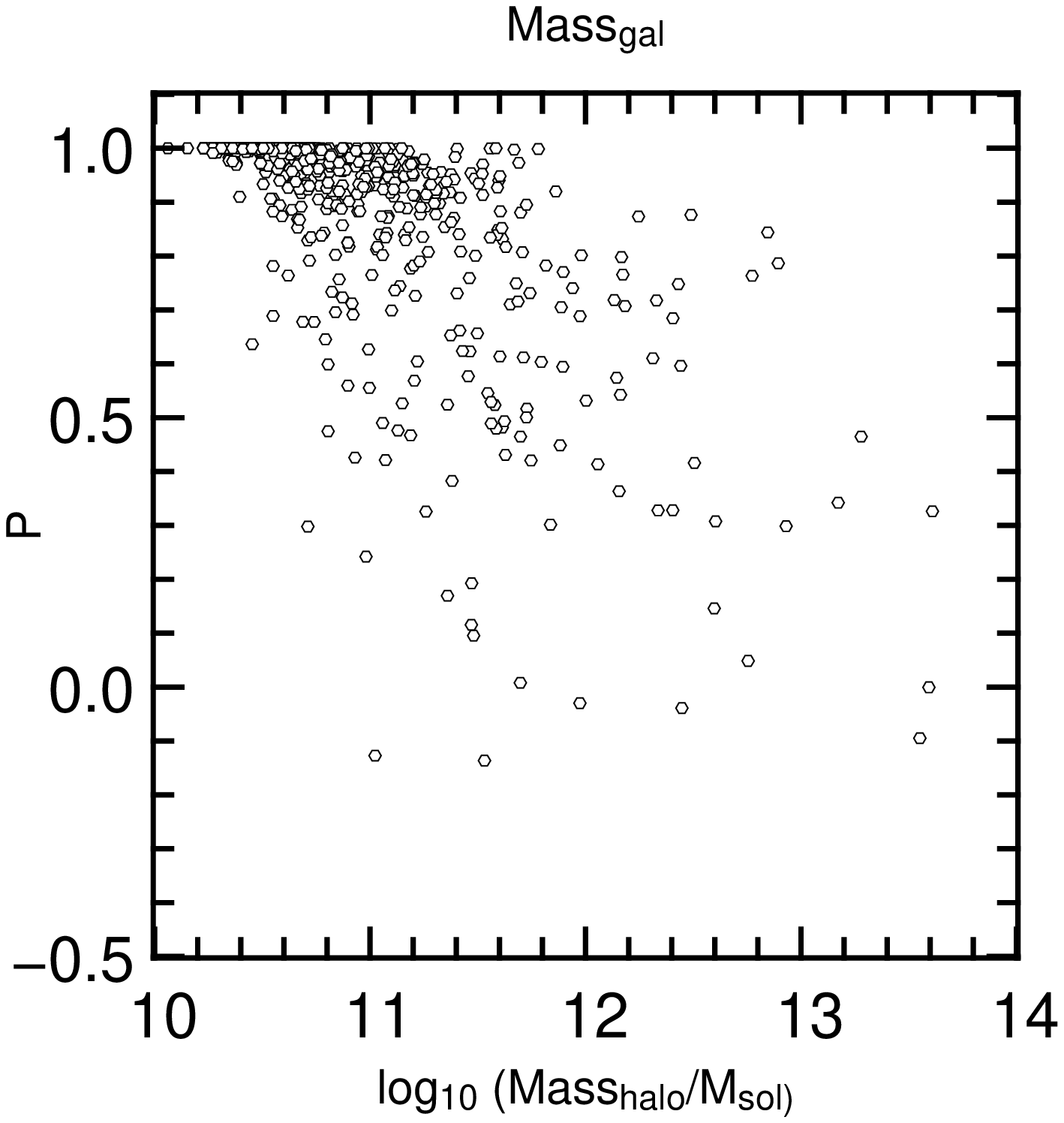}
\includegraphics[scale=0.33]{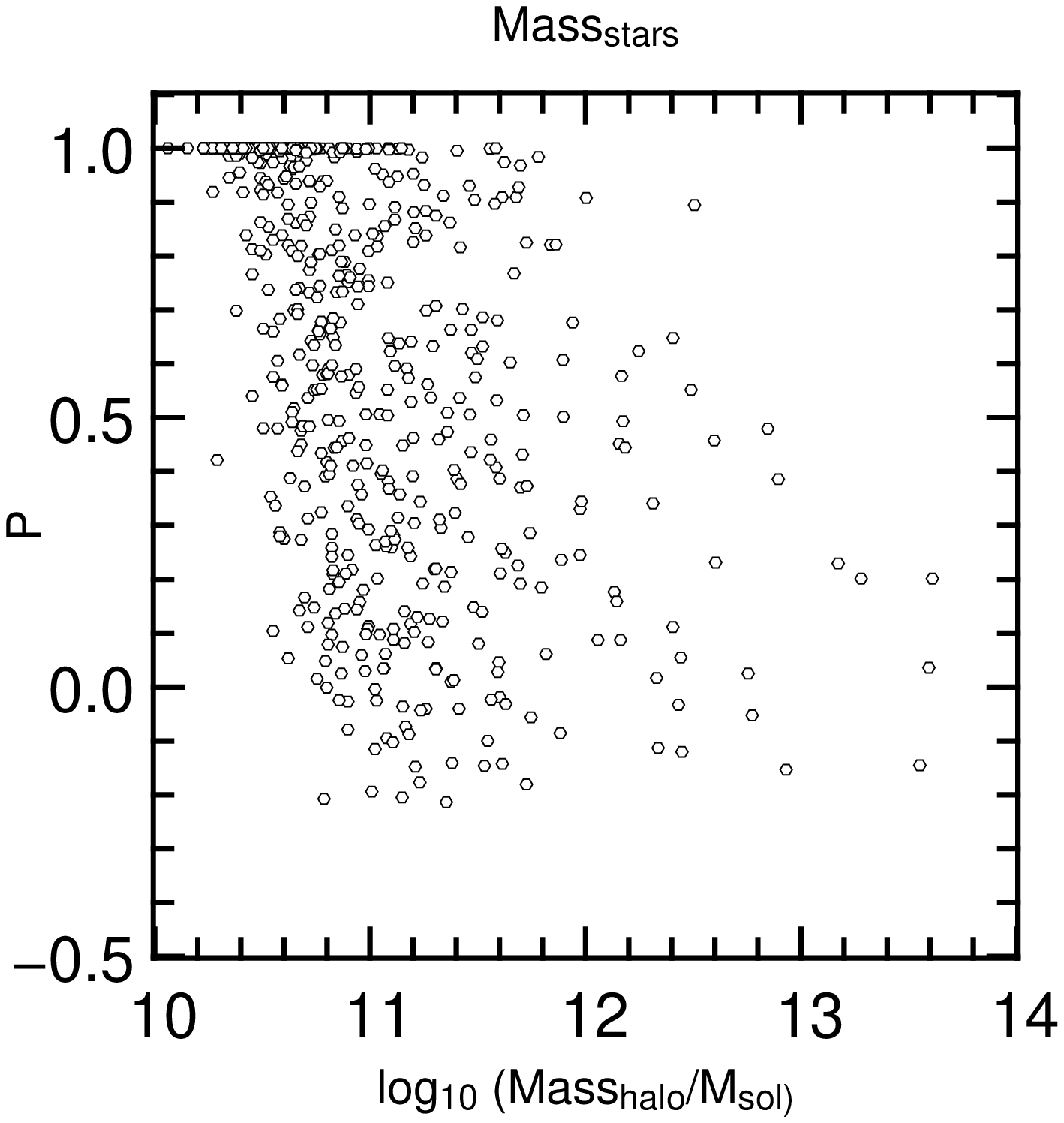}
\includegraphics[scale=0.33]{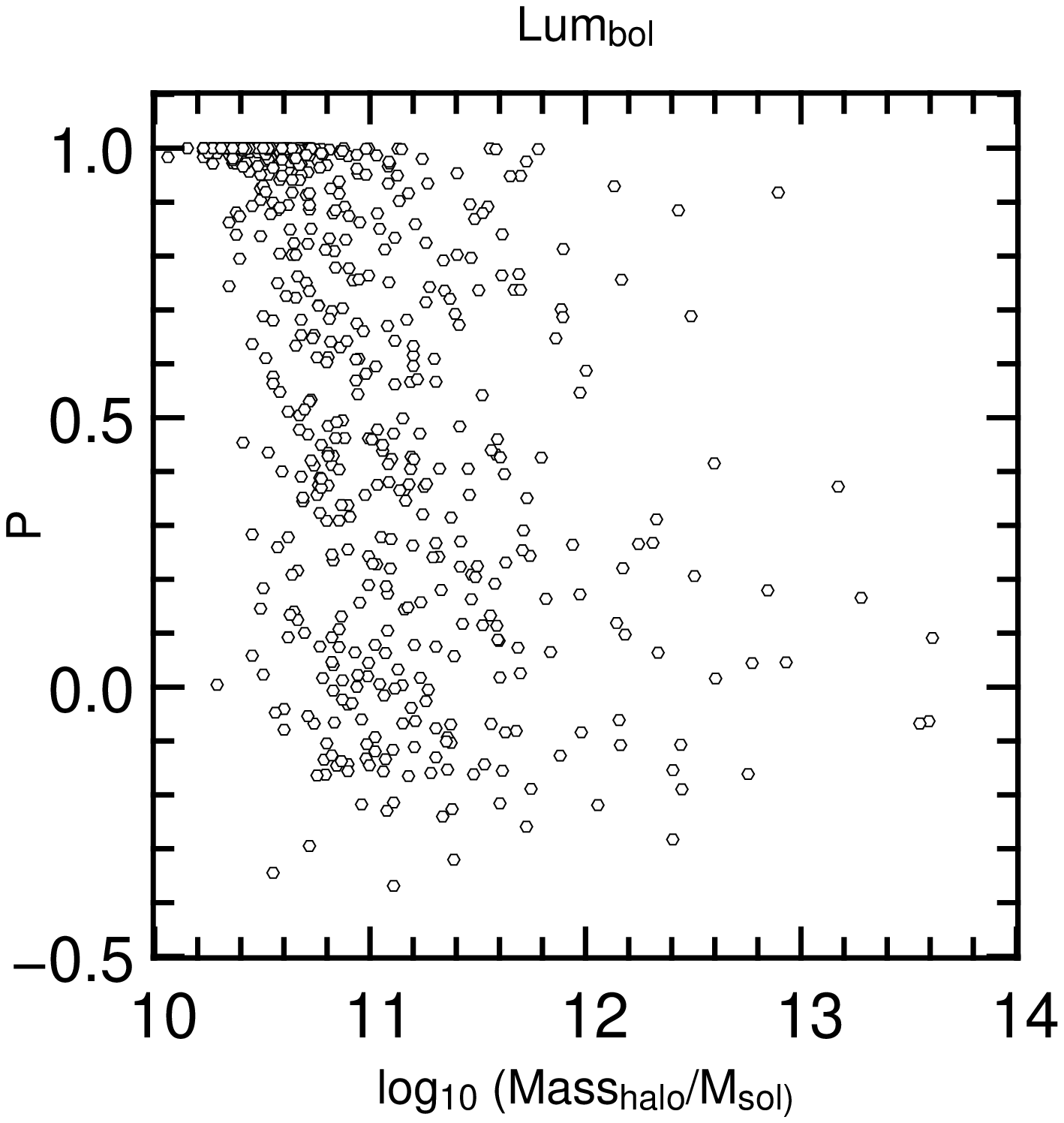}
\end{center}
\vspace{0.5cm}
\begin{center}
\includegraphics[scale=0.33]{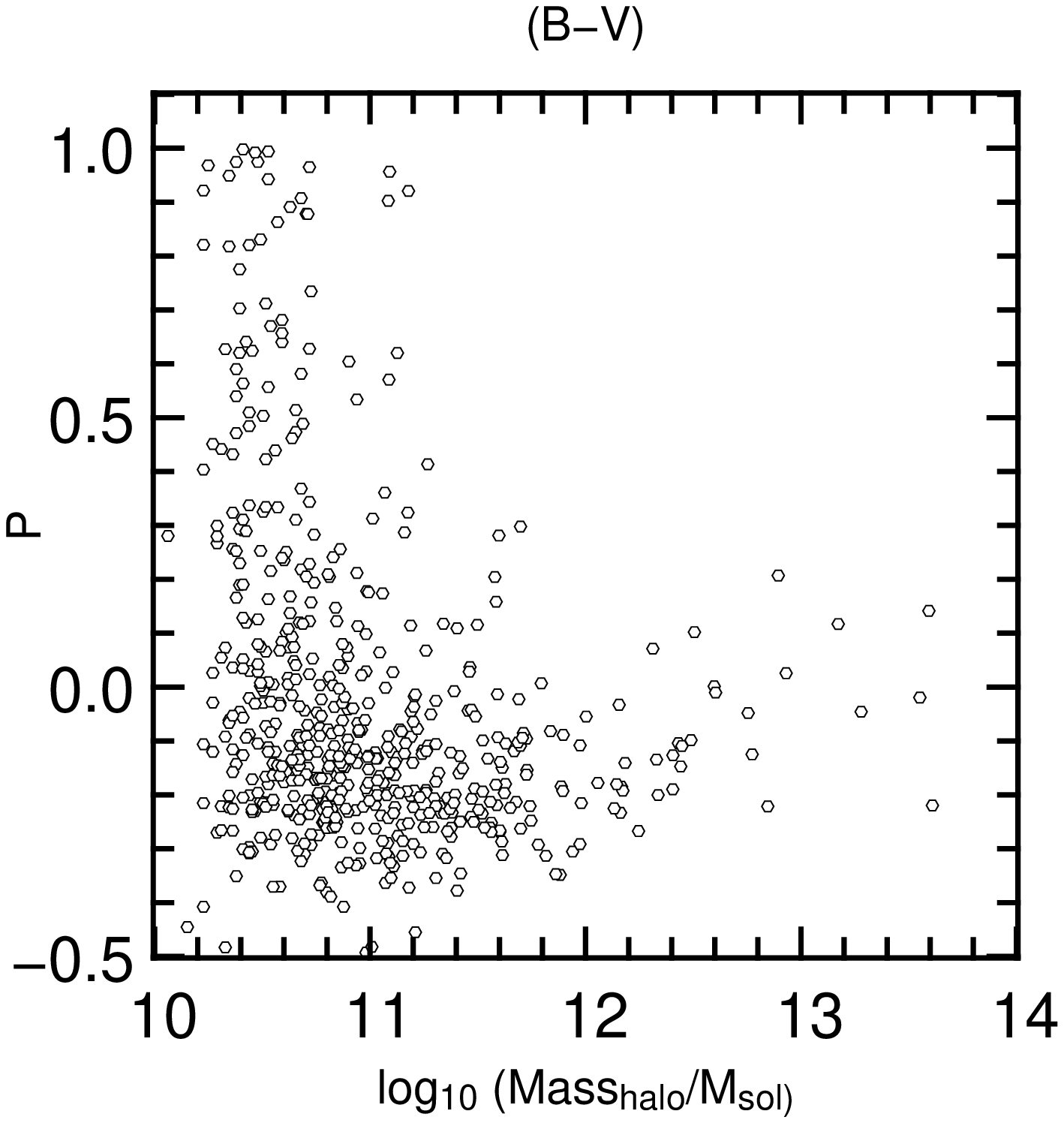}
\includegraphics[scale=0.33]{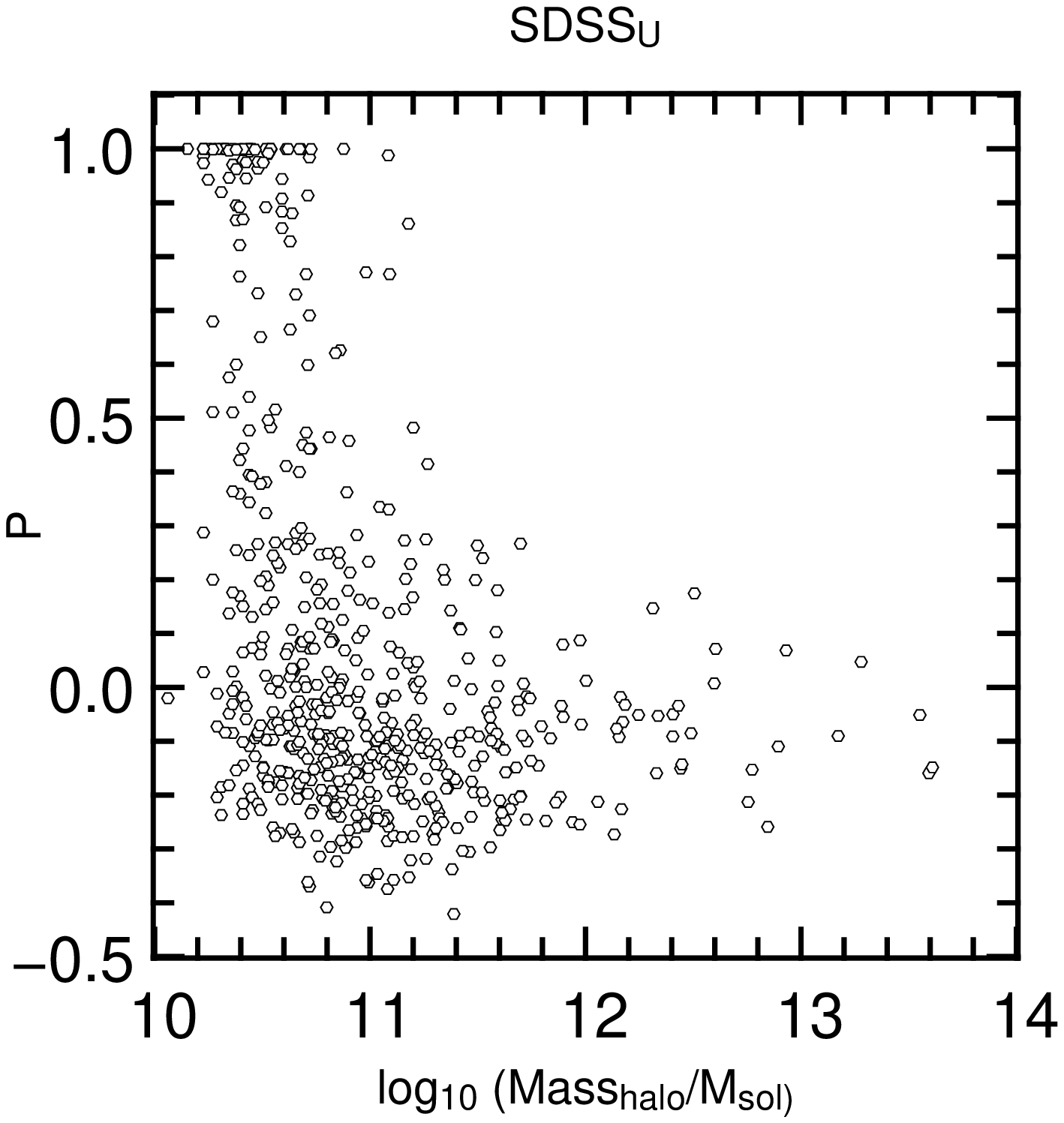}
\includegraphics[scale=0.33]{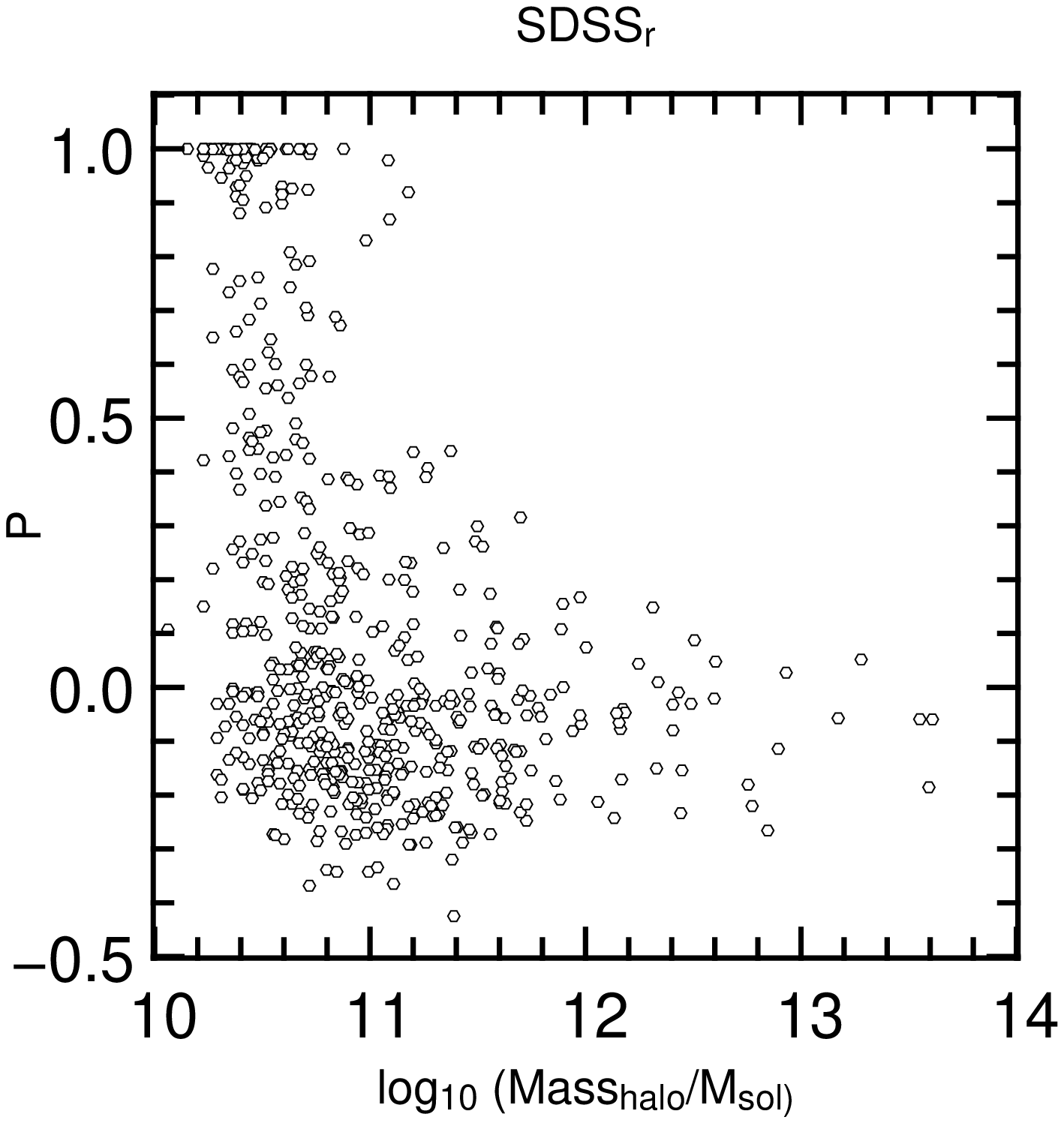}
\end{center}
\caption{\label{predictfigure} Results for the predictability of different
  properties. Each point represents a central galaxy. From left to right, top to bottom: total mass of the
  galaxy, stellar mass, bolometric luminosity, $(B-V)$ color, SDSS$_U$ and
  SDSS$_r$ magnitudes. The most dependent quantities on the star formation history
  (colors and magnitudes) are observed in general to have a very low
  predictability. The most predictable quantity respect to variations in the
  $\alpha$-$\epsilon$ plane is the total galactic mass.}
\end{figure*}

The predictability is bounded to $P\leq 1$. A value of $P\sim 1$
implies that the landscape is very smooth, while for values $P\leq 0$
the changes from neighboring sites can be high. 

We now turn to the results of the Fig.\ref{predictfigure}, where we plot the
predictability as a function of the logarithm of the mass of the host
halo, for all the galaxies in our study. Starting with the total
galaxy mass (stars and the gas) we can see that the galaxies
have high predictability, $P > 0.9$, in most of the cases. The situation is
quite different for the stellar mass and the bolometric luminosity. In these
cases the predictability ranges almost evenly between $0<P<1$, and we start seeing 
some fraction of points with negative predictability. In the case of the
$(B-V)$ colors and SDSS$_U$, SDSS$_r$ magnitudes we are in a totally different
ballpark as most of the landscapes have negative predictability, with a few
points over the range $0<P<1$.

The conclusion after these results is that we spot a landscape with a very
predictability $P<0.9$ we can be sure that the galaxy is sitting in halo less
massive than $3\times10^{11}$ M$_\odot$. In the same vein, when picking the
central galaxy in a halo of mass $> 10^{12}$ M$_\odot$, surely the predictability is
going to be lower $P<0.9$, or negative in the case of $(B-V)$ colors and
SDSS$_r$, SDSS$_U$ magnitudes.

\subsection{P-Weighted Landscape Variance}
\begin{figure*}
\begin{center}
\includegraphics[scale=0.33]{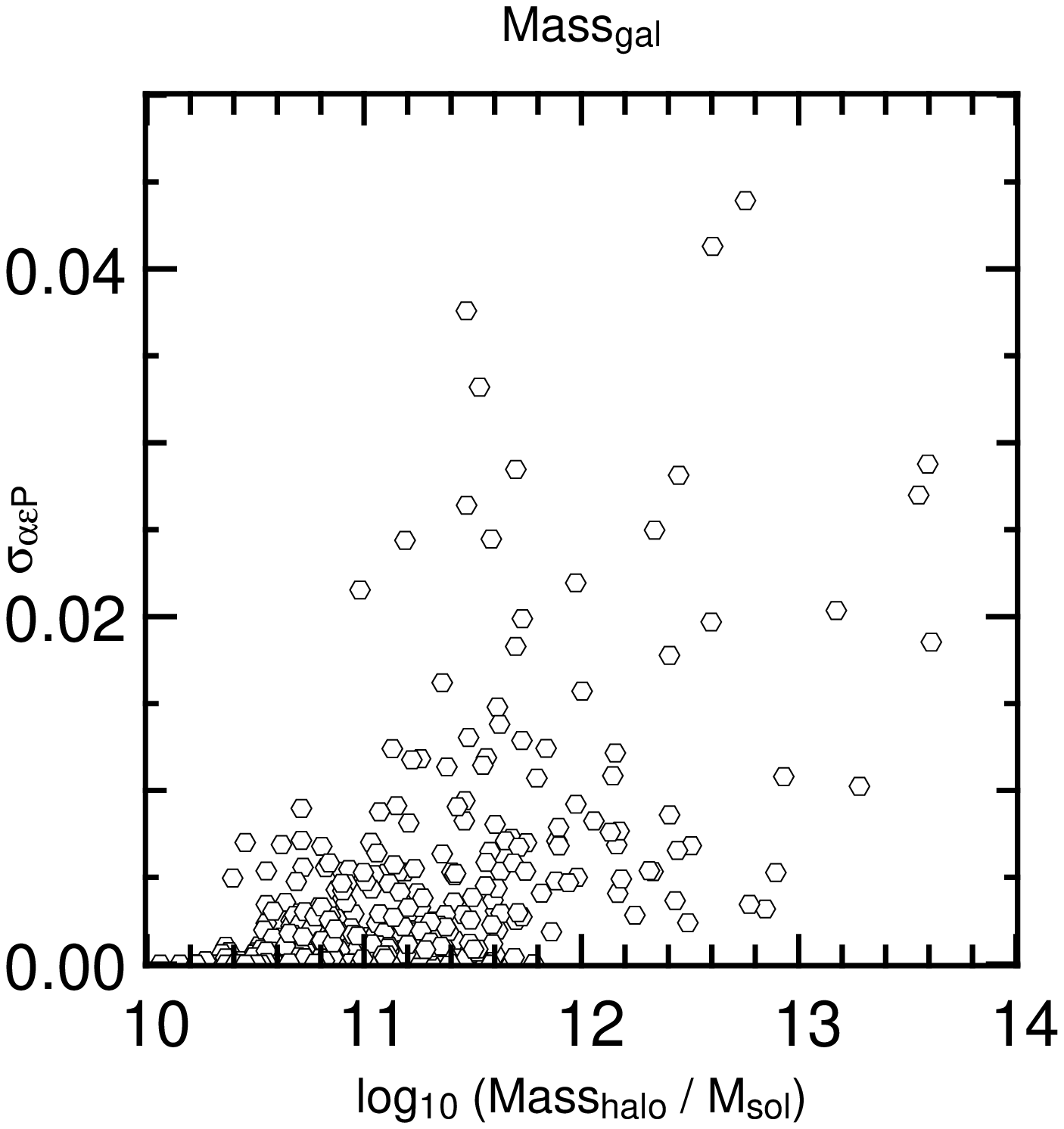}
\includegraphics[scale=0.33]{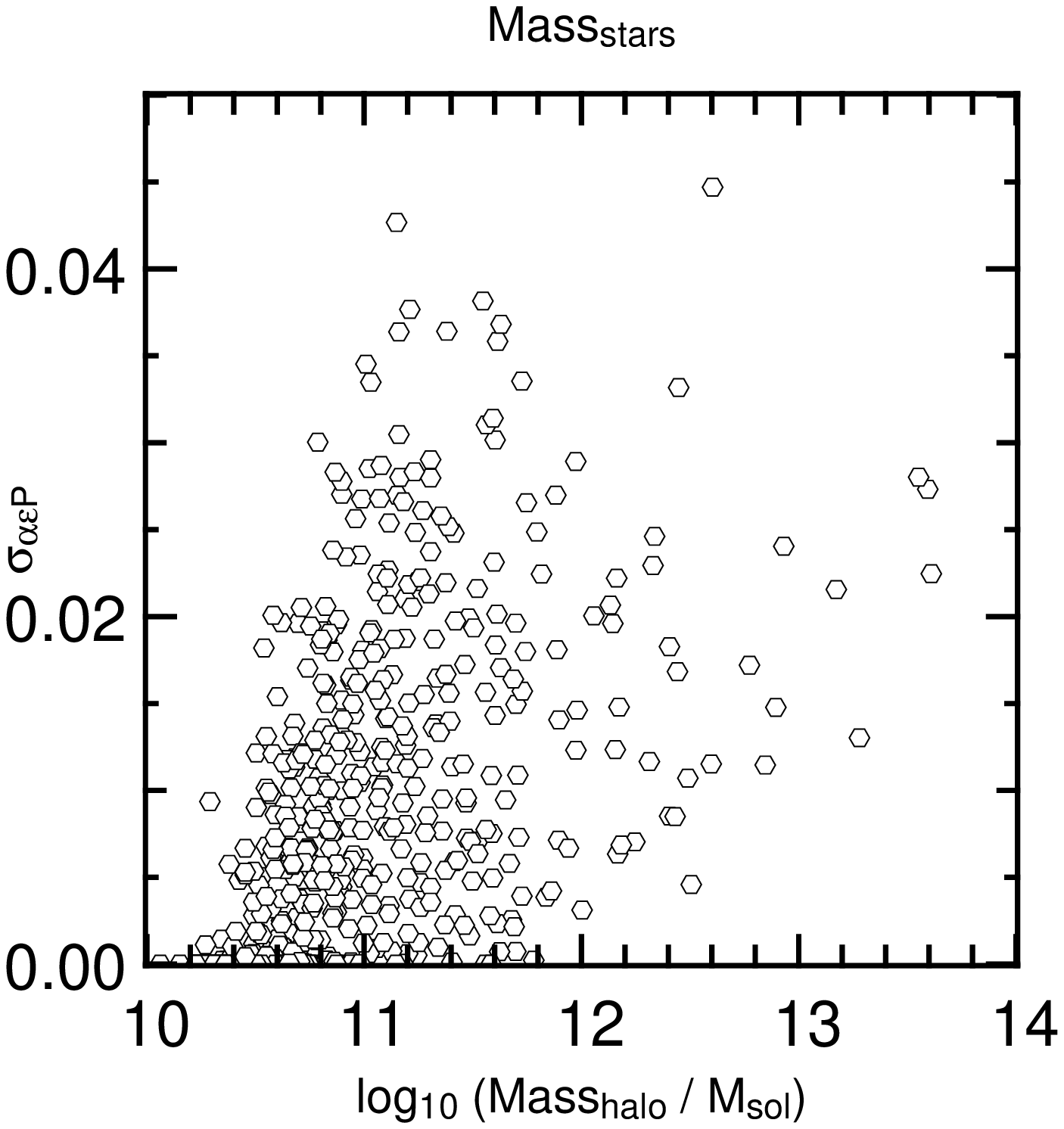}
\includegraphics[scale=0.33]{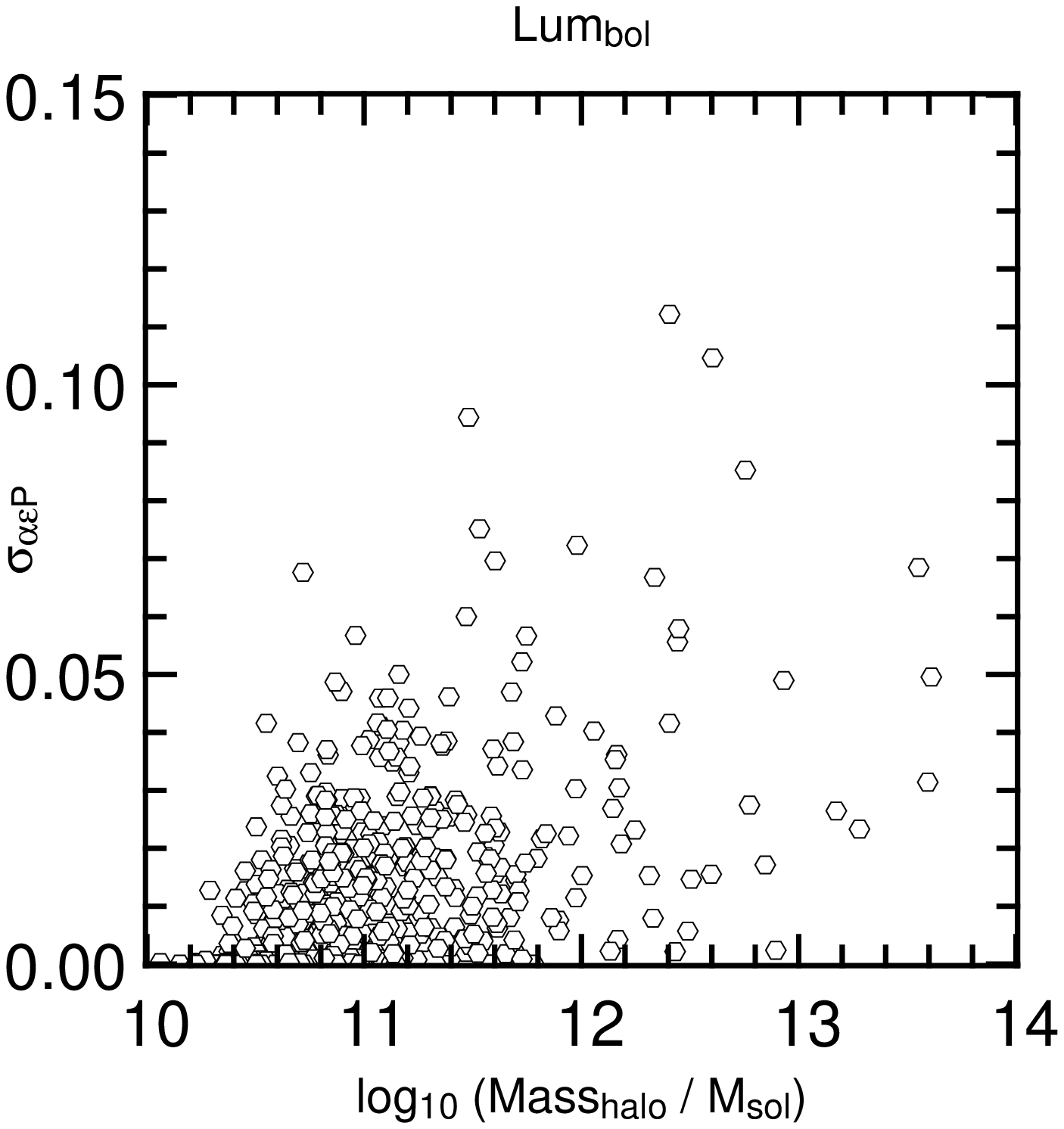}
\end{center}
\vspace{0.5cm}
\begin{center}
\includegraphics[scale=0.33]{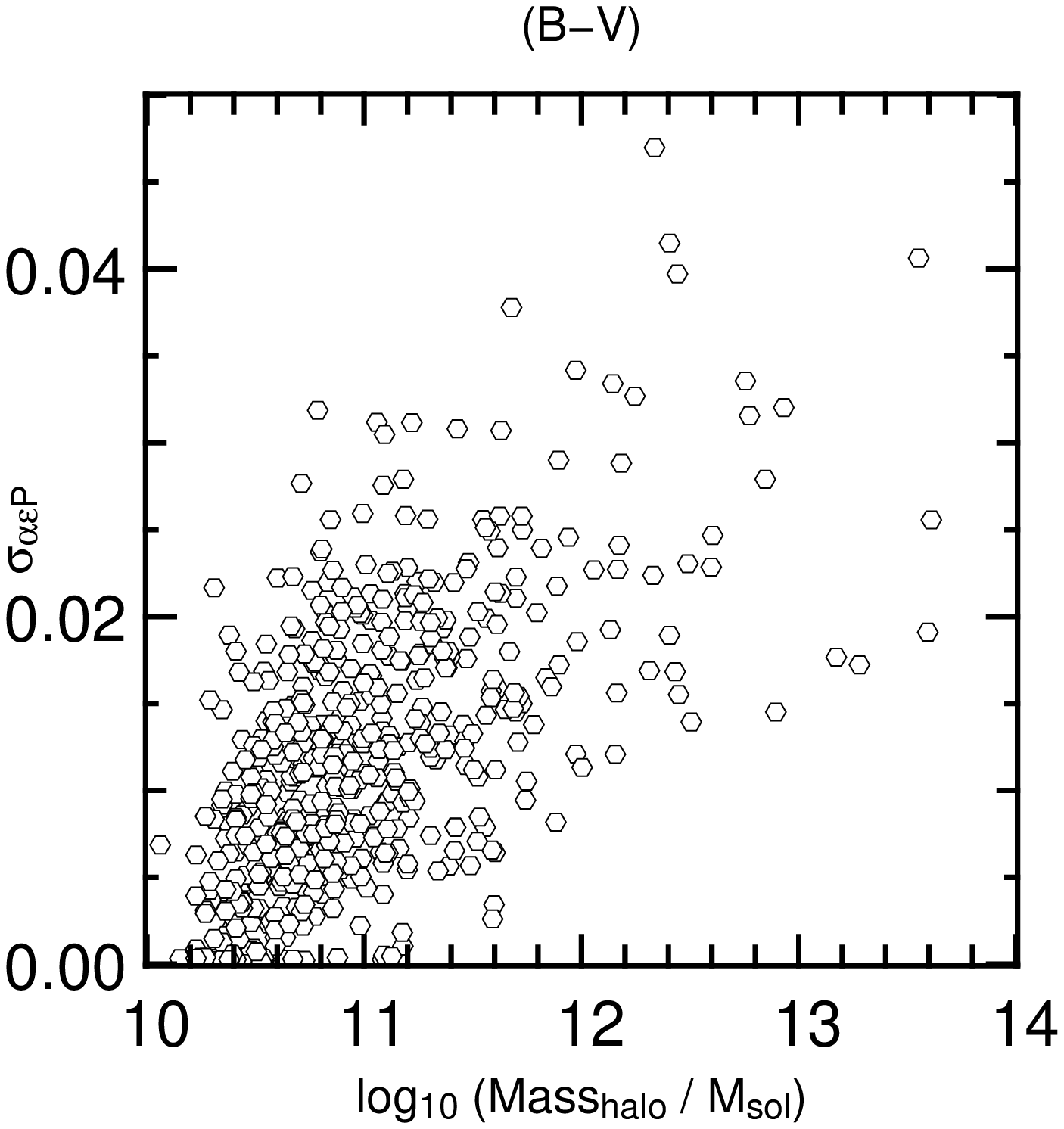}
\includegraphics[scale=0.33]{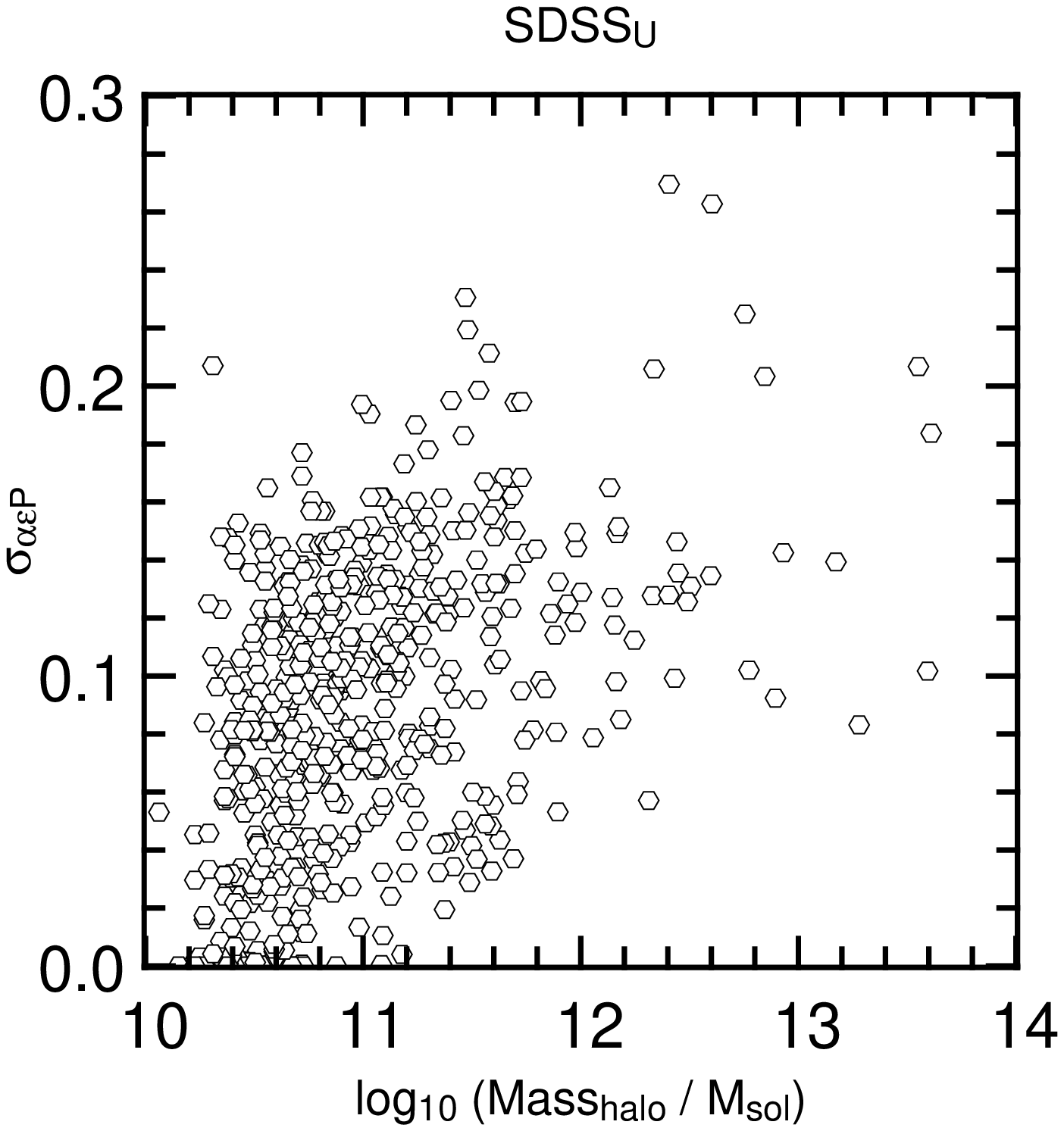}
\includegraphics[scale=0.33]{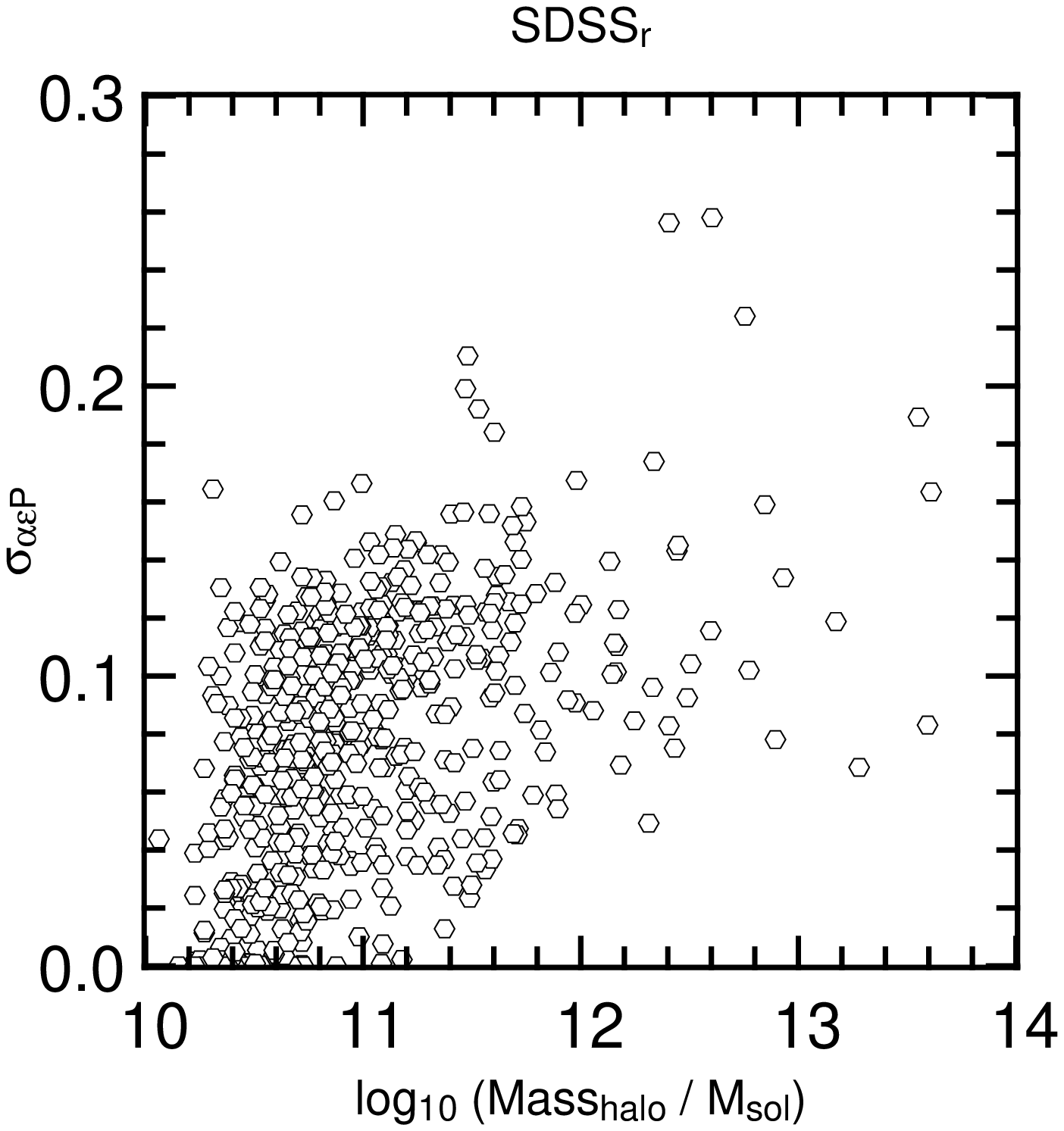}
\end{center}
\caption{\label{weightedfigure} Predictability-weighted variance
  ($\sigma_{\alpha\epsilon P}$) over the  $\alpha$-$\epsilon$ landscape
  (Eq. \ref{sigma_P}). This quantity can
  be used as an estimation of the expected change in a given galactic
  property when a variation $(\delta\alpha,\delta\epsilon)$ is
  performed on the input parameters. In the case of magnitudes it can
  be easily as high as $0.5$ for a $1\%$ variation in $(\alpha, \delta)$.
  A trend seems to indicate that this weighted variance increases with the dark
  matter halo mass.}
\end{figure*}

We explore now the second way of quantification of our results. It is
based on the landscape variance $\sigma_{\alpha\epsilon}$ over the the 320
points in the $\alpha$-$\epsilon$ plane (Eq.\ref{land_variance}) and the
predictability P (Eq.\ref{predigo}).   

We want to weight the landscape variance by the information obtained
through the predictability $P$. Performing a normalization in this way we can
have an idea about how much should be expected to vary a given galactic
property after performing a perturbation ($\delta \alpha, \delta\epsilon$). 

Using the variance $\sigma_{\alpha\epsilon}$ alone can be misleading in the case of
a high predictability landscape, because it could over-estimate the variation
of performing a ($\delta \alpha, \delta\epsilon$) perturbation. 

We propose then, the $P$-weighted variance

\begin{equation}
  \sigma_{\alpha\epsilon P}  = (1-e^{P-1})\times \sigma_{\alpha\epsilon},
\label{sigma_P}
\end{equation}

which has the property of being bound between $0\leq \sigma_{\alpha\epsilon
  P}\leq\sigma_{\alpha\epsilon}$ for the possible values of the predictability
$-\infty<P\leq 1$.

The general trend  (Fig.\ref{weightedfigure}) shows a growth in the
$P$-weighted predictability with halo mass, consistent with the fact that the
largest values for the predictability come along with large values for the
landscape variance. The results for the total mass landscape and the
bolometric luminosity stand apart, as this mass trend is less clear than for
the other galactic properties.

The values can be used to measure the variation one can expect from a
$1\%$ perturbation in $(\alpha,\epsilon)$. If we recall that this is calculated
from a $1$-$\sigma$ variance, the value of $\sigma_{\alpha\epsilon P}$ that
could effectively bracket the fluctuations over the $\alpha$-$\epsilon$ plane
would be three times that value. It means that for the most massive halos of
mass $\sim 10^{13}$ M$_{\odot}$ one can expect, at least, a variation of $\sim 0.5$ in
the SDSS$_U$ magnitudes or $\sim 0.1$ for  the $(B-V)$ colors after $1\%$
variation in $(\alpha,\epsilon)$. The variation in the total and stellar mass
could achieve at least $\sim 0.1$ dex.

\subsection{Cosmological Variance}

Now we compare the landscape variance $\sigma_{\alpha\epsilon}$ with the scale imposed by the cosmological context. We select from
the simulated cosmological box  (with $\alpha$ and $\epsilon$ in the
center of the $\alpha$-$\epsilon$ plane) all the dark matter halos with the
same mass (within $1\%$) as the parent halo of the galaxy under study. From
this halo population, we calculate the variance $\sigma_{halo}$
of the galactic properties of our interest for its central galaxies.

\begin{figure*}
\begin{center}
\includegraphics[scale=0.33]{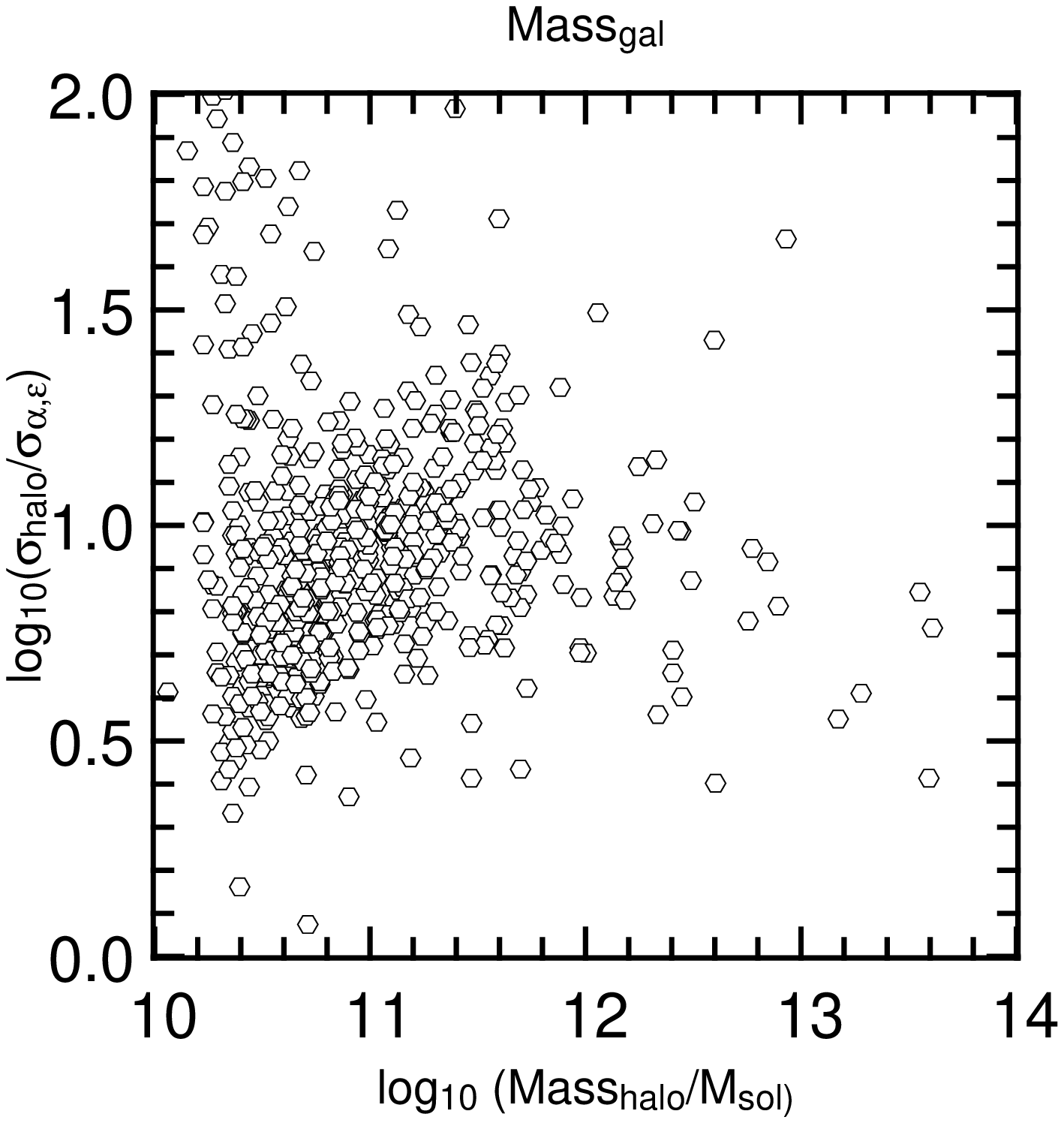}
\includegraphics[scale=0.33]{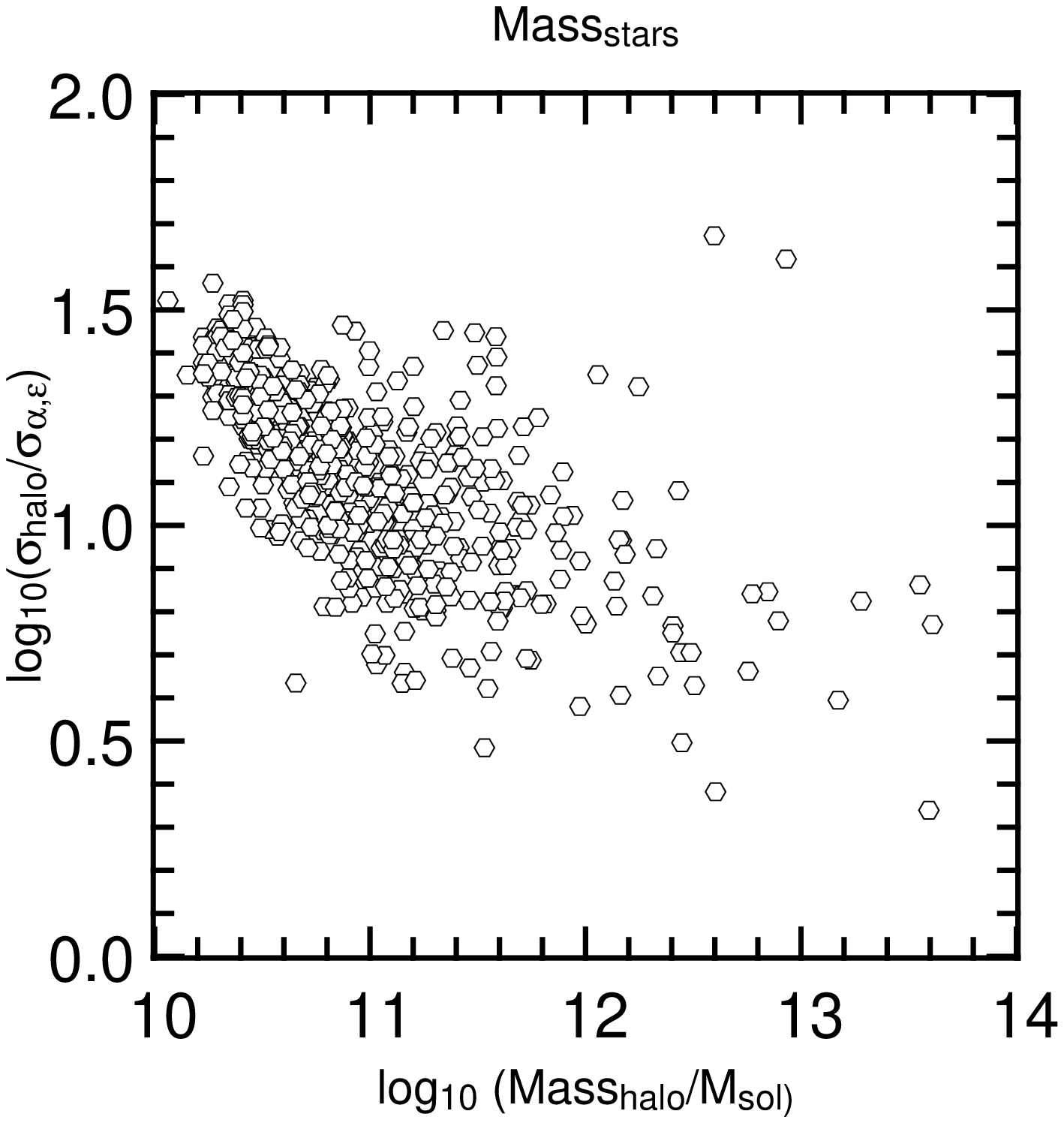}
\includegraphics[scale=0.33]{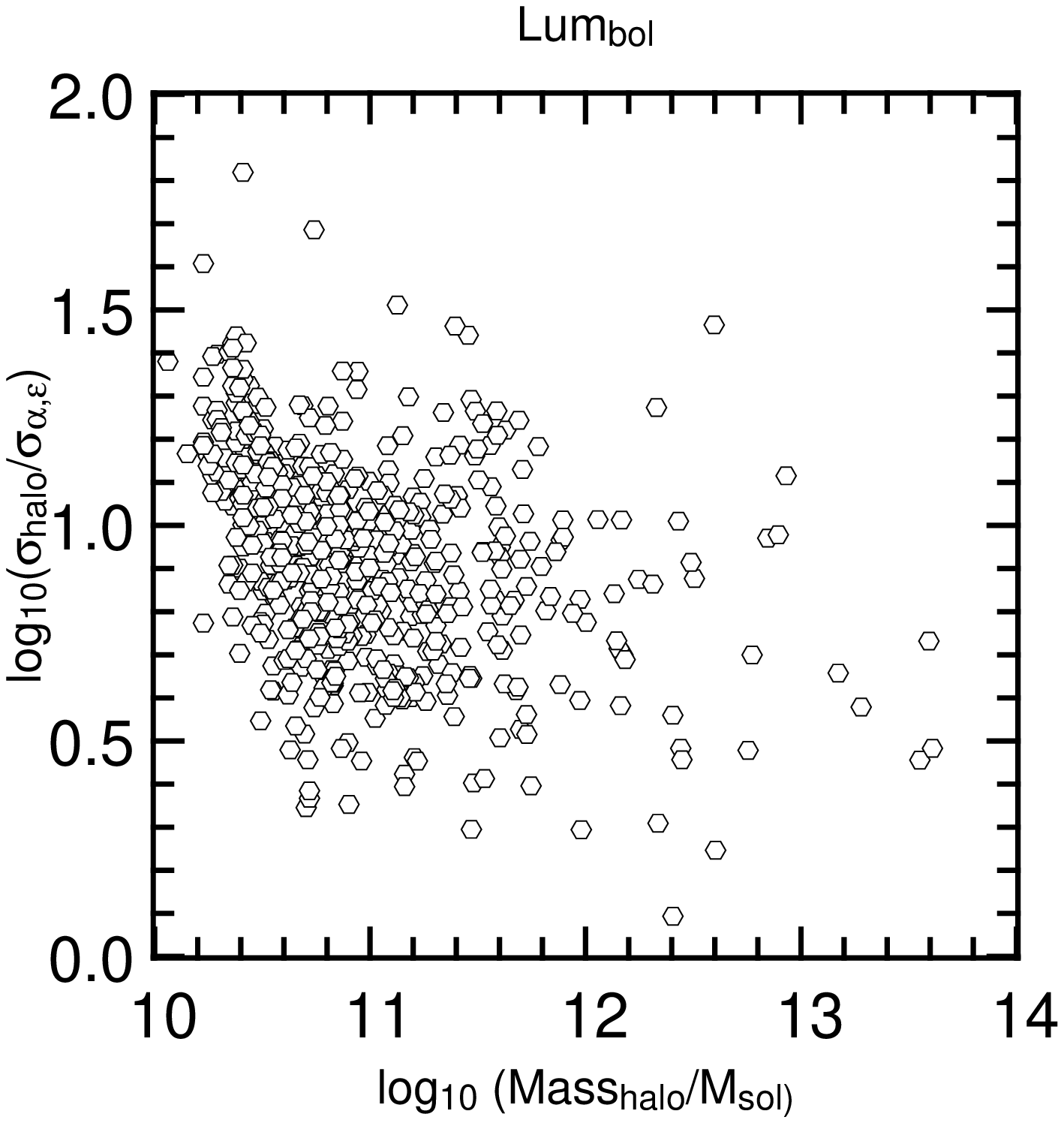}
\end{center}
\vspace{0.5cm}
\begin{center}
\includegraphics[scale=0.33]{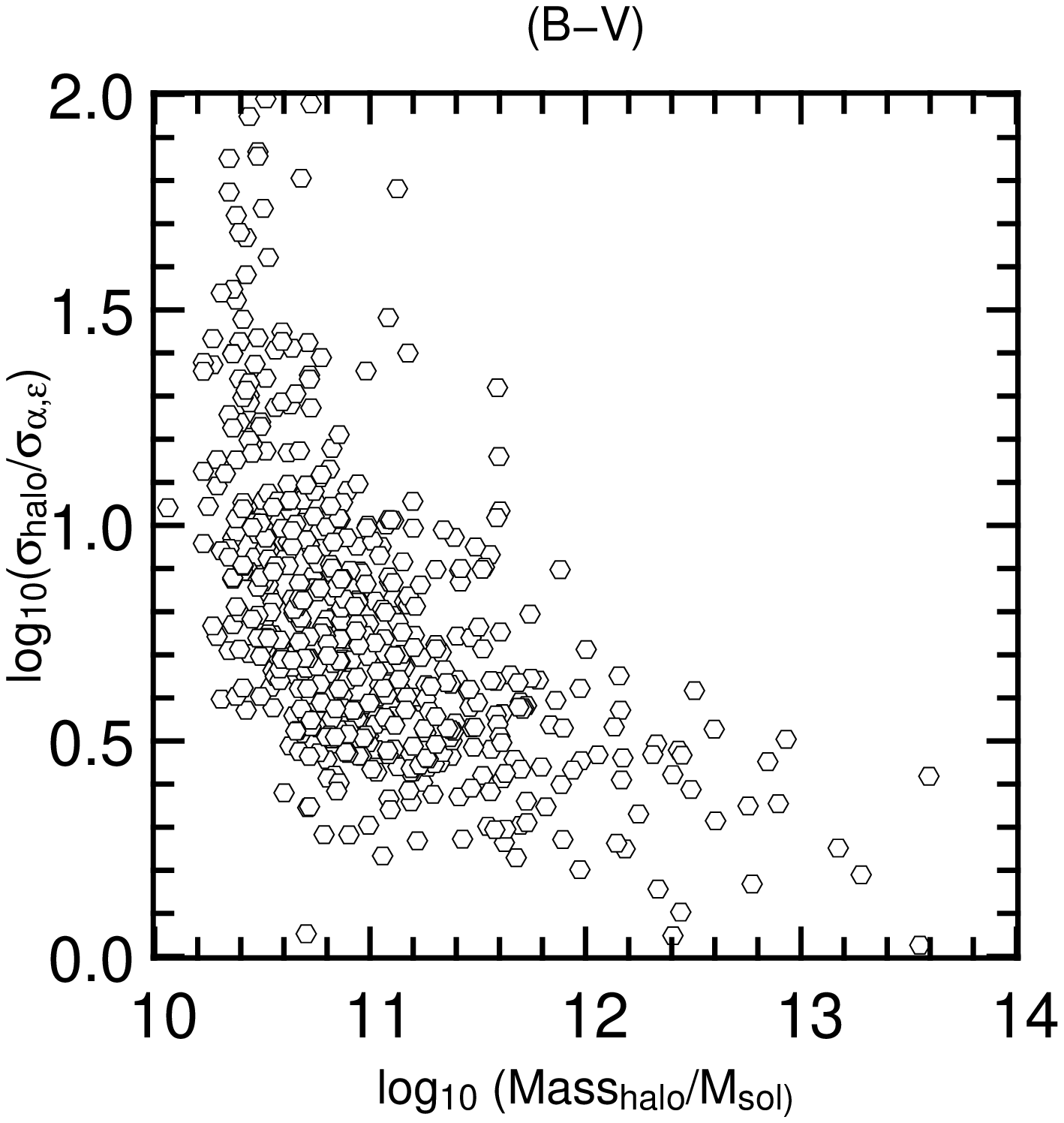}
\includegraphics[scale=0.33]{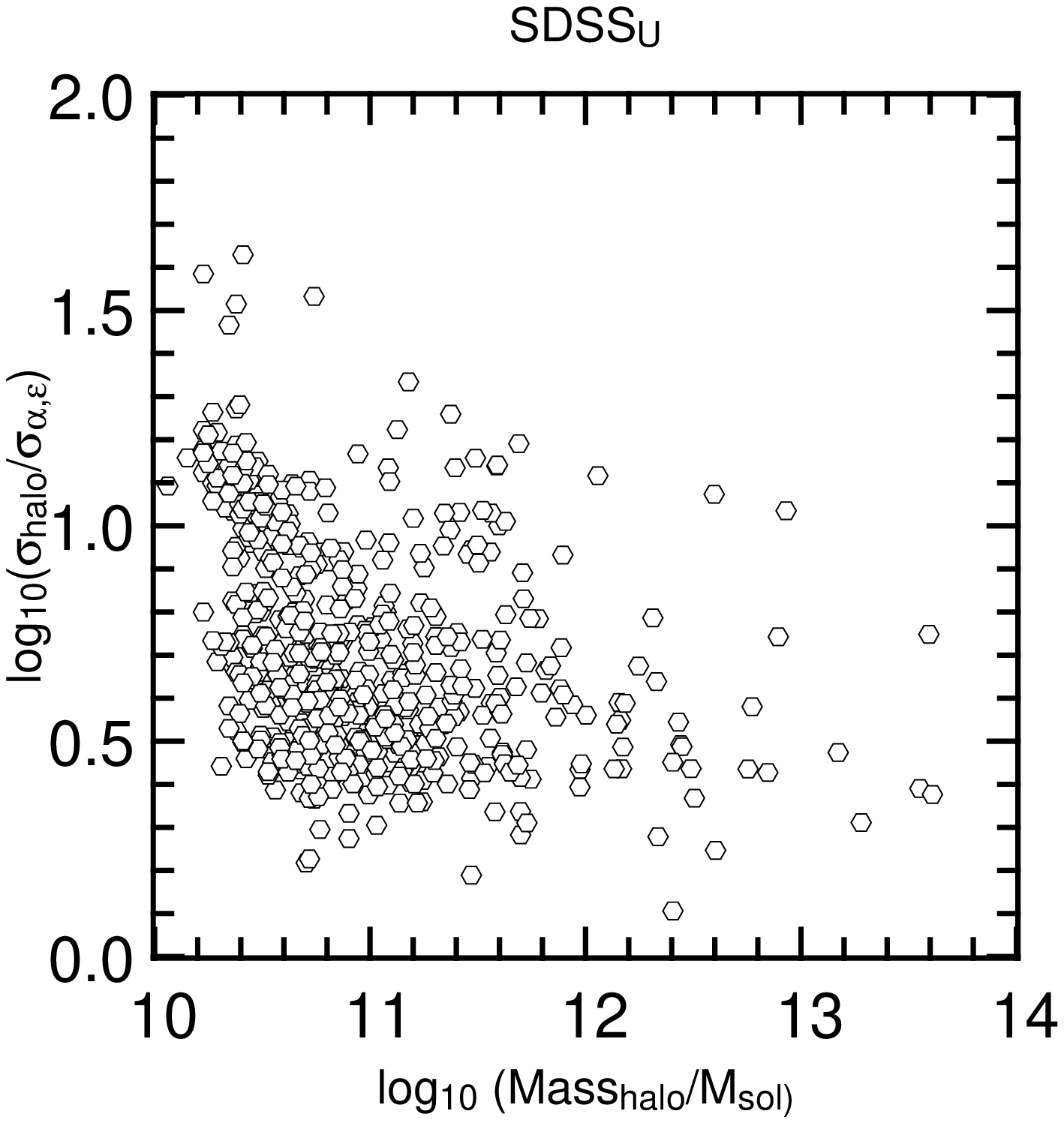}
\includegraphics[scale=0.33]{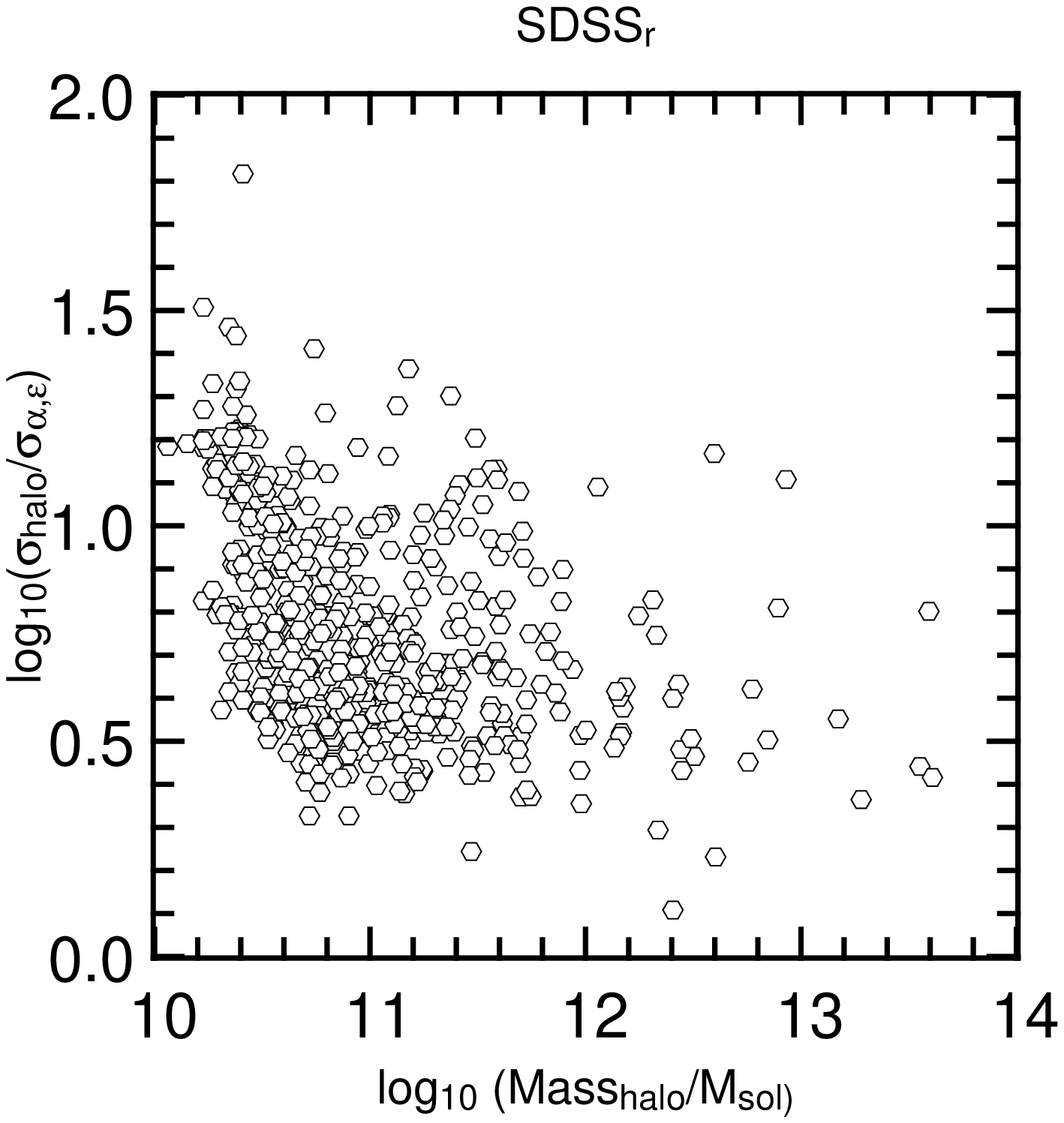}
\end{center}
\caption{\label{cosmofigure} Logarithm of the ratio of the variance calculated over a subset of halos in the simulated
  cosmological  box ($\sigma_{halo}$) to the landscape   variance $\sigma_{\alpha\epsilon}$.
  Except for the total galaxy mass, the general trend
  seems to indicate that the ratio is lower for higher halo masses. For
  magnitudes and colours, a variation in the $\alpha,\epsilon$ input
  parameters can be as important as a $1\%$ fluctuation in the halo mass. For
  lower masses, the halo mass fluctuations are dominant.} 
\end{figure*}

We show in Fig.\ref{cosmofigure} the logarithm of the ratio of the two variances
($\sigma_{halo}/\sigma_{\alpha\epsilon}$) as a function of halo
mass. For all the cases (except the total galactic mass) we observe the the
ratio $\sigma_{halo}/\sigma_{\alpha\epsilon}$ diminishes as the
halo mass grows. 

The case for the  $B-V$ color and the SDSS$_r$ and SDSS$_U$
band magnitude seems special.  For high masses the variance ratio is of the same order of
magnitude ($\sigma_{halo}/\sigma_{\alpha\epsilon}<3$,
$\log_{10}(\sigma_{halo}/\sigma_{\alpha\epsilon})<0.5$) in most of the cases.

This suggests that for central galaxies in low mass halos (where the predictability tends to be high)
the variance over its properties is dominated mostly by the possible mass of
the host halo. While for high masses (where the predictability tends to be low) the variance
for the galactic properties can be equally important if we vary the halo mass
or if we make a variation in the star formation and feedback efficiencies. 

\section{Discussion}

We have used $N$-body simulations and a semi-analytic model of galaxy
formation to explore the consequences of small perturbations in the input
parameters of the semi-analytic model. Specifically we varied the scalar parameters regulating the
star formation $\alpha$ and supernovae feedback efficiency $\epsilon$. We
followed some physical properties for $600$ central to gauge the effect of the perturbations.

This experiment was motivated both by the interest of performing a
description of a semi analytic models, making abstraction of the details in
the model , and test the significance of this approach to infer the
distinctive footprint of the different baryonic processes.

We find that depending on the halo mass, there are two kinds of
different of behavior respect to the change in the input parameters. There is a
smooth variation of the galactic properties for low mass halos, and a
seemingly random unpredictable behavior limited to high mass halos.   

We have quantified this behavior through an objective scalar function
called  predictability, $P$, defined in Eq.\ref{predigo}. Values $P\simeq 1$ mean
a rather smooth and predictable response,  while $P\leq 0$  point a  more
unpredictable behavior. The highest predictability is found only in low mass halos while
high mass halos present almost always negative values of $P$.  
This notion of predictability depends on the property we are looking
at. In our case the total galactic mass and the total stellar mass
showed the general trend of high predictability on the $\alpha
$-$\epsilon$ plane (upper panels in Fig.\ref{predictfigure}). While the $(B-V)$
color and the SDSS$_U$ and SDSS$_r$ bands showed almost always negative
predictabilities (lower panel in Fig.\ref{predictfigure}).  

Then, we computed the variance $\sigma_{\alpha\epsilon}$ over the
$\alpha$-$\epsilon$, $\sigma_{\alpha\epsilon}$ plane and weighted it by the
predictability over the same landscape. This helped us to estimate the
possible variation in a given property after performing a change
$\delta\alpha-\delta\epsilon$ in the input parameters. Using this measure we
found that for high mass halos on should expect rather large variations in the
galactic properties from a small perturbation in the baryonic parameters
(Fig.\ref{weightedfigure}). 

In order to give a scale to the landscape fluctuations for a given galaxy, we compare the
landscape variance with the variance in a subset of central galaxies of halos taken from the whole
cosmological box simulation. The halos are selected to have similar mass as
the parent halo of the galaxies we are studying . The general trend showed that
at higher halo masses the variance coming from the modulation of the $\alpha$
and $\epsilon$ parameters is on the same order of magnitude as the variation
on the galaxy properties over the whole box. The opposite trend is only found
for low halo masses  (Fig.\ref{cosmofigure}).   

In this particular case of perturbation related to star formation and
supernova feedback, it seems that the quantities that exhibit a dependence on the full star
formation history (magnitudes and colours) are the most sensitive to
variations of the $\alpha - \epsilon$ parameters. For instance, most of the trends we found
are not found for the total galaxy mass.

In general, all this evidence seems to point towards a picture where the central galaxies
hosted in massive halos, which have grown mainly through mergers, are
the most  sensitive to small variations of the baryonic parameters
in a way that is comparable of doing a significant variation on the mass of
the host halo.

From these results one can expect that if the variation of the
others scalar parameters in the model is performed, the predictability $P$,
landscape variance $\sigma_{\lambda_i}$ and $P$-weighted landscape variance
$\sigma_{\lambda_i P}$ should be higher than the values quoted in this paper. 
It is very unlikely that the variations of other parameters could cancel
 out exactly the influence of the efficiencies $\alpha$ $\epsilon$.

There is a hierarchy of causes for this behavior that must be
explored. First of all, might this be a signature of the hierarchical
build of galaxies? In such a picture it is easy to imagine that mild
perturbations at early times might add up to finally yield very
different values for very similar initial conditions. This would
account for the relatively large values of the variance over the
$\alpha$-$\epsilon$ plane compared to the intrinsic variances over the
whole population of similar halos in the cosmological volume. But
probably not for the low predictability values.

Could this be an artifact coming from the semi-analytic models of
galaxy formation? In these models, generally the distinction between what is to
be considered as the central galaxy depend on which galaxy is the most
massive. This is ambiguous when various galaxies inside
a dark matter halo have similar masses, in that case the selection of
the central galaxy might be subject to noise. This could explain in
part the seemingly random landscapes for high mass haloes.

On the last level of the hierarchy, could this be coming from
\emph{our code}? This is impossible to confirm without performing the
same kind  of experiment with another fully fledged semi-analytic
model. Which take us to the issue of comparison between semi-analytic
models of galaxy formation. The predictability, as a meaningful scalar
objective function, opens the possibility to measure the biases from different
semi-analytic codes. This could allow the comparison
of different codes based on its numerical performance, going beyond the
rather ill-posed strategy of comparison based on astrophysical performance,
i.e. reproducing observations.

Finally, the small perturbations we made on the scalar parameters were
constructed  to not have any effect on the mean properties of the
galaxies such as  the luminosity function. It means that formally
the galaxies we have produced at every perturbation are consistent
with the overall galaxy population. 

As a consequence of all this, studies making use of selected subpopulations from 
a wider population generated using semi-analytic models, should bear in
mind that this smaller population might not be unique. The dispersion
on this subsample of galaxies, coming from the perturbations that can
be induced on every parameter in the model, should be explicitly
stated. Including that dispersion (in the form of error bars, for instance) seems
a necessary condition to make a \emph{fair use} of semi-analytic models,
acknowledging in an explicit manner its limitations on predictability.

\section*{Acknowledgments}
Many thanks to  Dylan Tweed, Julien Devriendt, J\'er\'emy Blaizot, Bruno Guiderdoni
and Christophe Pichon for their hard work around the version of the GalICS Code
used in this paper. This work was performed in the framework of the HORIZON
Project (France).


\begin{thebibliography}{}

\bibitem[\protect\citeauthoryear{{Abel}, {Bryan} \& {Norman}}{{Abel}
  et~al.}{2002}]{2002Sci...295...93A}
{Abel} T.,  {Bryan} G.~L.,    {Norman} M.~L.,  2002, Science, 295, 93

\bibitem[\protect\citeauthoryear{{Baugh}}{{Baugh}}{2006}]{2006RPPh...69.3101B}
{Baugh} C.~M.,  2006, Reports of Progress in Physics, 69, 3101

\bibitem[\protect\citeauthoryear{{Bell}, {Baugh}, {Cole}, {Frenk} \&
  {Lacey}}{{Bell} et~al.}{2003}]{2003MNRAS.343..367B}
{Bell} E.~F.,  {Baugh} C.~M.,  {Cole} S.,  {Frenk} C.~S.,    {Lacey} C.~G.,
  2003, \mnras, 343, 367

\bibitem[\protect\citeauthoryear{{Croton}, {Springel}, {White}, {De Lucia},
  {Frenk}, {Gao}, {Jenkins}, {Kauffmann}, {Navarro} \& {Yoshida}}{{Croton}
  et~al.}{2006}]{2006MNRAS.365...11C}
{Croton} D.~J.,  {Springel} V.,  {White} S.~D.~M.,  {De Lucia} G.,  {Frenk}
  C.~S.,  {Gao} L.,  {Jenkins} A.,  {Kauffmann} G.,  {Navarro} J.~F.,
  {Yoshida} N.,  2006, \mnras, 365, 11

\bibitem[\protect\citeauthoryear{{Davis}, {Efstathiou}, {Frenk} \&
  {White}}{{Davis} et~al.}{1985}]{FOF}
{Davis} M.,  {Efstathiou} G.,  {Frenk} C.~S.,    {White} S.~D.~M.,  1985, \apj,
  292, 371

\bibitem[\protect\citeauthoryear{{De Lucia} \& {Blaizot}}{{De Lucia} \&
  {Blaizot}}{2007}]{2007MNRAS.375....2D}
{De Lucia} G.,  {Blaizot} J.,  2007, \mnras, 375, 2

\bibitem[\protect\citeauthoryear{{Gottl{\"o}ber}, {Yepes}, {Khalatyan},
  {Sevilla} \& {Turchaninov}}{{Gottl{\"o}ber}
  et~al.}{2006}]{2006AIPC..878....3G}
{Gottl{\"o}ber} S.,  {Yepes} G.,  {Khalatyan} A.,  {Sevilla} R.,
  {Turchaninov} V.,  2006, in {Manoz} C.,  {Yepes} G.,  eds, The Dark Side of
  the Universe Vol.~878 of American Institute of Physics Conference Series,
  {Dark and baryonic matter in the MareNostrum Universe}.
pp~3--9

\bibitem[\protect\citeauthoryear{{Hatton}, {Devriendt}, {Ninin}, {Bouchet},
  {Guiderdoni} \& {Vibert}}{{Hatton} et~al.}{2003}]{galicsI}
{Hatton} S.,  {Devriendt} J.~E.~G.,  {Ninin} S.,  {Bouchet} F.~R.,
  {Guiderdoni} B.,    {Vibert} D.,  2003, \mnras, 343, 75

\bibitem[\protect\citeauthoryear{{Hayashi} \& {White}}{{Hayashi} \&
  {White}}{2007}]{2007arXiv0709.3933H}
{Hayashi} E.,  {White} S.~D.~M.,  2007, ArXiv e-prints, 709

\bibitem[\protect\citeauthoryear{{Kennicutt}
  Jr.}{{Kennicutt}}{1998}]{1998ApJ...498..541K}
{Kennicutt} Jr. R.~C.,  1998, \apj, 498, 541

\bibitem[\protect\citeauthoryear{{Monaco}, {Fontanot} \& {Taffoni}}{{Monaco}
  et~al.}{2007}]{MORGANA}
{Monaco} P.,  {Fontanot} F.,    {Taffoni} G.,  2007, \mnras, 375, 1189

\bibitem[\protect\citeauthoryear{{Norman}, {Bryan}, {Harkness}, {Bordner},
  {Reynolds}, {O'Shea} \& {Wagner}}{{Norman}
  et~al.}{2007}]{2007arXiv0705.1556N}
{Norman} M.~L.,  {Bryan} G.~L.,  {Harkness} R.,  {Bordner} J.,  {Reynolds} D.,
  {O'Shea} B.,    {Wagner} R.,  2007, ArXiv e-prints, 705

\bibitem[\protect\citeauthoryear{{Pascual} \& {Levin}}{{Pascual} \&
  {Levin}}{1999}]{predictpaper}
{Pascual} M.,  {Levin} S.~A.,  1999, Ecology, 80, 2225

\bibitem[\protect\citeauthoryear{{Silk}}{{Silk}}{2001}]{2001MNRAS.324..313S}
{Silk} J.,  2001, \mnras, 324, 313

\bibitem[\protect\citeauthoryear{{Somerville} \& {Kolatt}}{{Somerville} \&
  {Kolatt}}{1999}]{1999MNRAS.305....1S}
{Somerville} R.~S.,  {Kolatt} T.~S.,  1999, \mnras, 305, 1

\bibitem[\protect\citeauthoryear{{Somerville} \& {Primack}}{{Somerville} \&
  {Primack}}{1999}]{1999MNRAS.310.1087S}
{Somerville} R.~S.,  {Primack} J.~R.,  1999, \mnras, 310, 1087

\bibitem[\protect\citeauthoryear{{Spergel}, {Verde}, {Peiris}, {Komatsu},
  {Nolta}, {Bennett}, {Halpern}, {Hinshaw}, {Jarosik}, {Kogut}, {Limon},
  {Meyer}, {Page}, {Tucker}, {Weiland}, {Wollack} \& {Wright}}{{Spergel}
  et~al.}{2003}]{2003ApJS..148..175S}
{Spergel} D.~N.,  {Verde} L.,  {Peiris} H.~V.,  {Komatsu} E.,  {Nolta} M.~R.,
  {Bennett} C.~L.,  {Halpern} M.,  {Hinshaw} G.,  {Jarosik} N.,  {Kogut} A.,
  {Limon} M.,  {Meyer} S.~S.,  {Page} L.,  {Tucker} G.~S.,  {Weiland} J.~L.,
  {Wollack} E.,    {Wright} E.~L.,  2003, \apjs, 148, 175

\bibitem[\protect\citeauthoryear{{Springel}, {Frenk} \& {White}}{{Springel}
  et~al.}{2006}]{LSS_SFW}
{Springel} V.,  {Frenk} C.~S.,    {White} S.~D.~M.,  2006, \nat, 440, 1137

\bibitem[\protect\citeauthoryear{{Taffoni}, {Monaco} \& {Theuns}}{{Taffoni}
  et~al.}{2002}]{2002MNRAS.333..623T}
{Taffoni} G.,  {Monaco} P.,    {Theuns} T.,  2002, \mnras, 333, 623

\end{thebibliography}
\end{document}